\documentclass[
aps,
pra,
superscriptaddress,
showpacs,
showkeys,
notitlepage,
twocolumn, 
numbers,
longbibliography,
nofootinbib]
{revtex4-1}

\usepackage[T1]{fontenc} 
\usepackage[utf8]{inputenc} 
\usepackage[english]{babel} 
\usepackage{graphicx, xcolor}
\usepackage{amsfonts, amssymb, amsmath, mathrsfs, calc, array, mathtools}
\usepackage[separate-uncertainty=true]{siunitx}
\usepackage{hyperref}
\usepackage{abraces}

\usepackage{tikz}
\usetikzlibrary{quantikz}

\def\bra#1{\mathinner{\langle{#1}|}}
\def\ket#1{\mathinner{|{#1}\rangle}}

\def\ontop#1#2{\setbox0\hbox{#2}\copy0\llap{\raise\ht0\hbox{#1}}}

\newcommand{\defeq}{\vcentcolon=}

\definecolor{darkblue}{rgb}{0,0,0.93} 
\definecolor{darkred}{rgb}{0.8,0,0} 
\definecolor{darkgreen}{rgb}{0,0.7,0}

\hypersetup{
colorlinks=true, 
linkcolor=darkblue, 
citecolor=darkred, 
filecolor=darkblue, 
urlcolor=darkblue
}

\def\NOTEPA#1{{\textcolor{red}{\bf [#1]}}}  
\begin{document}

\title{Quantum circuits for discrete-time quantum walks \\ with position-dependent coin operator}

\author{Ugo Nzongani}
\email{ugo.nzongani@universite-paris-saclay.fr}
\affiliation{Universit{\'e} Paris-Saclay, CNRS, ENS Paris-Saclay, INRIA, Laboratoire M{\'e}thodes Formelles, 91190 Gif-sur-Yvette, France}

\author{Julien Zylberman}
\email{julien.zylberman@obspm.fr}
\affiliation{{Sorbonne Université, Observatoire de Paris, Université PSL, CNRS, LERMA, 75005 Paris, France}}

\author{Carlo-Elia Doncecchi}
\affiliation{Universit{\'e} Paris-Saclay, CNRS, ENS Paris-Saclay, INRIA, Laboratoire M{\'e}thodes Formelles, 91190 Gif-sur-Yvette, France}





\author{Armando P{\'e}rez}
\email{armando.perez@uv.es}
\affiliation{Departamento de F{\'i}sica Te{\'o}rica and IFIC, Universidad de Valencia and CSIC, Dr.\ Moliner 50, 46100 Burjassot, Spain}

\author{Fabrice Debbasch}
\affiliation{{Sorbonne Université, Observatoire de Paris, Université PSL, CNRS, LERMA, 75005 Paris, France}}

\author{Pablo Arnault}
\email{pablo.arnault@inria.fr}
\affiliation{Universit{\'e} Paris-Saclay, CNRS, ENS Paris-Saclay, INRIA, Laboratoire M{\'e}thodes Formelles, 91190 Gif-sur-Yvette, France}




\begin{abstract}
The aim of this paper is to build quantum circuits that implement discrete-time quantum walks having an arbitrary position-dependent coin operator.
The position of the walker is encoded in base $2$: with $n$ wires, each corresponding to one qubit, we encode $2^n$ position states.
The data necessary to define an arbitrary position-dependent coin operator is therefore exponential in $n$.
Hence, the exponentiality will necessarily appear somewhere in our circuits.

We first propose a circuit implementing the position-dependent coin operator, that is naive, in the sense that it has exponential depth and implements sequentially all appropriate position-dependent coin operators.
We then propose a circuit that ``transfers'' all the depth into ancillae, yielding a final depth that is linear in $n$ at the cost of an exponential number of ancillae.
The main idea of this \emph{linear-depth circuit} is to implement in parallel all coin operators at the different positions.
Reducing the depth exponentially at the cost of having an exponential number of ancillae is a goal which has already been achieved for the problem of loading classical data on a quantum circuit \cite{APPS2021} (notice that such a circuit can be used to load the initial state of the walker).
Here, we achieve this goal for the problem of applying a position-dependent coin operator in a discrete-time quantum walk.

Finally, we extend the result of Ref.\ \cite{WGMAG2014} from position-dependent unitaries which are diagonal in the position basis to position-dependent \emph{$2\times 2$-block-diagonal} unitaries: indeed,  we show that for a position dependence of the coin operator (the block-diagonal unitary) which is smooth enough, one can find an efficient quantum-circuit implementation approximating the coin operator up to an error $\epsilon$ (in terms of the spectral norm), the depth and size of which scale as $O(1/\epsilon)$.

A typical application of the efficient implementation would be the quantum simulation of a relativistic spin-1/2 particle on a lattice, coupled to a smooth external gauge field;
notice that recently, quantum spatial-search schemes have been developed which use gauge fields as the oracle, to mark the vertex to be found \cite{ZD2021, FZAD2022}.
A typical application of the linear-depth circuit would be when there is spatial noise on the coin operator (and hence a non-smooth dependence in the position).

\end{abstract}

\keywords{}

\pacs{}

\maketitle


\section{Introduction}

Quantum walks (QWs) are models of quantum transport on graphs, for example on the hypercube for algorithmic applications \cite{Moore2002, Potoek2009}, or on lattices for both algorithmic (spatial search, see below) and physical applications (simulation of physical situations of interest, see below).
They are a universal model of computation \cite{Childs2009, CGW13, Lovett2010}, which enjoys in certain situations exponential speedup with respect to classical random walks \cite{Childs2003, Kempe2003b}, and has been used to conceive various algorithms, for example of element distinctness \cite{Amb07a}, or of search problems \cite{Childs2004, Tulsi2008, MNRS11, Ambainis2013, FG2014, RGAM20}, for which the speedup is polynomial with respect to classical search.

QWs exist in continuous and in discrete time \cite{Kempe2003a}.
Continuous-time QWs (CQWs) are described by a local Hamiltonian defined on some graph, while discrete-time QWs (DQWs) are defined by a unitary evolution operator between two consecutive time instants which is strictly local.
Various connections have been made between CQWs and DQWs \cite{Strauch06b, Strauch07a, ChildsCD2009, PP16, Schmitz2016, APAF18, DMA19}.

QWs have also been used extensively for the simulation of various physical theories and phenomena:
this application is obvious in the case of CQWs given their definition mentioned above; in the case of DQWs, their use as simulators of physical theories often relies on the fact that many DQWs have as continuum limit the Dirac equation \cite{book_FH65, Arrighi_higher_dim_2014}.
In this paper, we focus on DQWs.
DQWs can thus simulate spin-1/2 particles in external Abelian \cite{DDMEF12a, AD16a, AD16b} and non-Abelian \cite{AMBD16} Yang-Mills gauge fields, as well as in relativistic gravitational fields \cite{DMD13b, DMD14, AD17, Arrighi_curved_1D_15, AF17}.
Other various phenomena have been exhibited \cite{DMP2016, Bru2016b, MMDMplus2017, APP20}.
Lattices other than square ones have also been used for DQWs, such as triangular, hexagonal \cite{ADMplus2018, GDW2019}, and cylindrical \cite{JAD21} ones.
Action principles for such DQWs have been proposed \cite{Debbasch2019a, AC2022}, and connections to lattice gauge theories have been made \cite{MMAMP18, APAF18, CGWW18, AC2022}.
Also, the question of a lattice relativistic covariance for DQWs has been investigated \cite{AFF14a, Debbasch2019b}.
Multiparticle DQWs are known as quantum cellular automata \cite{Arrighi2019, Farrelly2020}, and they have been used to build certain simple models of lattice gauge theory with strict relativistic lightcone at the lattice level, such as free theories \cite{DAP16}, the bosonic $\phi^4$ theory in any dimension \cite{FS2020}, or quantum electrodynamics in $1+1$ \cite{ABF20, SADM2022}, $1+2$ and $1+3$ dimensions \cite{EDMMplus22}.

QWs have been implemented using different setups \cite{book_Manouchehri}, such as (i) photons in various optical devices \cite{Trompeter06, Schreiber10a, Peruzzo10a, Kitagawa2012}, among which integrated optical fibers \cite{Sansoni11a, CORGplus2013, BNSO2017}, (ii) atoms trapped in arrays of light \cite{GASW13}, (iii) ion traps \cite{BCMS2019}, and (iv) superconductiong qubits \cite{KSBKplus2020}.
QWs have shown to be strongly demanding when it comes to their implementation on quantum computers; in fact, only a few steps can be reliably executed, before noise effects completely jeopardize the result \cite{AAMKplus2020, Shakeel2020a, GEZ2021, PPT2021}.
Various works studying the effect of noise on DQWs exist \cite{Ken07a, DiMolfetta2016, AMACplus20}.
During the NISQ era \cite{Preskill2018}, an efficient design of the quantum circuits implementing the desired QWs, is therefore crucial in order to make use of their potential as quantum simulators.
As said above, in this paper we concentrate on the DQW, although quantum circuits to implement continuous-time quantum walks (CQWs) have also been proposed and realized on different setups \cite{DW2009, QLMAplus2016, Ming2021}.

There are two reasons why we focus on DQWs and do not treat CQWs in this work.
First, DQWs can be used to simulate the Dirac equation \emph{while preserving the strict locality of the transport}, and it is often in that perspective that we are interested in DQWs.
Indeed, from one instant to the next one, a DQW transports the probability amplitude at some position towards adjacent positions, \emph{within a bounded neighborhood around the original position}, and this creates, at the discrete level, an equivalent of the continuum lightcone of relativistic physics, which is exhibited in particular by the Dirac equation.
In contrast, CQWs are nothing but tight-binding-like Hamiltonians on lattices, in continuous time, and as such they do not exhibit a strict lightcone as DQWs, they only have an approximate lightcone via Lieb-Robinson bounds.
An example of such a CQW is the Kogut-Susskind Hamiltonian, which is a discretization of the Dirac Hamiltonian.
Again, such CQW discretizations are structurally fundamentally different from DQW discretizations, since they do not respect the strict locality of the continuum Dirac equation.
The second reason why we focus on DQWs and do not treat CQWs in this work is the following.
DQWs are governed (in addition to the kind of lattice on which they are evolving), by a fundamental and versatile piece: the coin operator.
This coin operator can be tailored at will, at each time step, thus allowing the DQW to reproduce a large variety of phenomena, such as couplings of the matter field (the walker) to electromagnetic fields, gravitational fields, or such as localization effects arising from an external potential.
In particular, in the case of gravitational couplings, the point is that the spatial and temporal variations of the metric are encoded in the coin operator solely, on a fixed regular lattice, i.e., with no need of considering a curved lattice.
For CQWs, simulating such a phenomenon may imply a specific design of the graph geometry, which, again, is not needed with DQWs.
We do not exclude the possibility of engineering appropriate CQWs but, again, the general point we want to make is that, to the best of our knowledge, such a degree of plasticity in simulations has not been achieved by CQWs.

In order to realize DQWs on quantum computers, the first task is to efficiently translate the DQW into a quantum circuit; indeed, the programs of quantum computers are designed within the quantum circuit model, so that the Hilbert space is that of $n$ entangled qubits, whereas the Hilbert space of a DQW is that of a graph tensor the coin space.
With $n+1$ wires, one can encode $2^n=N$ positions, and the last wire is for the coin state.
The DQW is made of two operations:  one operation implements the coin-dependent shift, and the other implements the coin rotation; we recall these facts in Sec.\ \ref{sec:1}.
For the quantum-circuit implementation of the coin-dependent shift operation, see Sec.\ \ref{sec:shift}, we choose the Fourier-transform scheme proposed in Ref.\ \cite{Shakeel2020a}, recalled in Sec.\ \ref{subsec:QFT_scheme}, which demands less resources than the increment and decrement scheme of Ref.\ \cite{FOBH2005}.
Now, to date, all works proposing quantum circuits implementing DQWs, choose a uniform coin operator, i.e., a coin operator which does not depend on the position of the walker.
However, many applications of DQWs to physical simulations need to introduce a position-dependent coin operator, see above and in Sec.\ \ref{sec:conclusions}.
The purpose of this work is to design quantum circuits for the implementation of coin operators depending arbitrarily on the position of the walker.
Since there is one different coin operator per lattice position, the number of parameters of a position-dependent coin operator is linear in $N=2^n$, i.e., exponential in $n$.
This is why the depth of the first circuit we present, see Sec.\ \ref{sec:naive}, which we call naive circuit, is exponential in $n$; in that circuit, the coin operators for the different positions are applied sequentially, one after the other, hence the depth exponential dependence in $n$.
Then, in Sec.\ \ref{sec:linear}, we ``transfer'' all the depth of the naive circuit into ancillae, which yields a circuit the depth of which is, this time, linear in $n$, at the cost of introducing an exponential number of ancillae.
Let us mention that this aim of transferring all the exponential depth of a circuit into ancillae has already been achieved for the problem of loading classical data on a quantum circuit, i.e., quantum state preparation \cite{APPS2021, APLOplus2021}, and that such algorithms can be used to load the initial condition of our DQW.
Finally, in Sec.\ \ref{sec:efficient}, we  extend the result of Ref.\ \cite{WGMAG2014} from position-dependent unitaries which are diagonal in the position basis to position-dependent \emph{$2\times 2$-block-diagonal} unitaries.
Indeed,  we show that for a position dependence of the coin operator (the block-diagonal unitary) which is smooth enough, one can find an approximate quantum-circuit implementation (of this coin operator), the depth and size of which scale as $O(1/\epsilon)$, where $\epsilon$ is the error committed in the approximation in terms of the spectral norm of the coin operator; this $O(1/\epsilon)$ reaches efficiency according to the standard definition, for which efficiency means that the depth and size of the circuit scale no worse than $O(\text{poly}(n,1/\epsilon))$ \cite{Childs2004thesis}.
This efficient circuit for smooth position-dependent coin operators is likely to be usable for all DQWs coupled to gauge fields mentioned earlier in this introduction; notice that recently, quantum spatial-search schemes have been developed which use gauge fields as the oracle, to mark the vertex to be found \cite{ZD2021, FZAD2022}.
For spatial noise on the coefficients of the coin operator, the smoothness condition will not hold anymore and one can use the linear-depth circuit mentioned earlier in this introduction.
In Sec.\ \ref{sec:Implementation}, we implement the naive, the linear-depth, and the Walsh circuits, on the classical simulator of IBM's quantum processors, to give an idea of the results one obtains for a small number of position qubits.
In Sec.\ \ref{sec:conclusions}, we conclude and discuss our results.

\section{Discrete-time quantum walk on the circle, with position-dependent coin}
\label{sec:1}

\subsection{The Hilbert space and the state of the system}
\label{subsec:Hilbert}

Consider a circle with $N$ nodes, labelled from $0$ to $N-1$.
It will become clear in Sec.\ \ref{subsec:encoding} why the discussion is simplified if we assume that $N$ is a power of $2$: we thus assume $N=2^n$ for some $n\in \mathbb{N}$.
To each node $k=0,...,N-1$ we associate a position quantum state $\ket{k}$, and we consider the position Hilbert space $\mathcal{H}_{\text{pos.}}$ spanned by the family $(\ket{k})_{k=0,...,N-1}$.
In addition to this position degree of freedom, our quantum state has an additional, ``internal'' degree of freedom, that we call coin: the coin state $\ket{c}$ belongs to a two-dimensional Hilbert space $\mathcal{H}_{0}$, a basis of which is $(\ket{\uparrow}, \ket{\downarrow})$.
Unless there is some possible ambiguity with the $\ket{k}$'s, we will actually use the notation $\ket{\uparrow} \equiv \ket{0} \defeq (1,0)^{\top}$ and $\ket{\downarrow} \equiv \ket{1}\defeq (0,1)^{\top}$, where $\top$ denotes the transposition, in particular since further down $\ket{c}$ will be associated to one wire of a quantum circuit.
Our quantum state thus belongs to the following total Hilbert space, $\mathcal{H} \defeq \mathcal{H}_{\text{pos.}} \otimes \mathcal{H}_{0}$, and can generically be written, at time $j$, as
\begin{equation}
\label{eq:state}
\ket{\psi_j} \defeq \sum_{k=0}^{N-1} \left( \psi^{\uparrow}_{j,k} \ket{k}\ket{0} +  \psi^{\downarrow}_{j,k} \ket{k}\ket{1} \right) \, .
\end{equation}

\subsection{The walk}

We consider, as the dynamics for our quantum state, Eq.\ \eqref{eq:state}, one standard form of a DQW:
\begin{equation}
\label{eq:the_walk_scheme}
\ket{\psi_{j+1}} = W^{(n)} \ket{\psi_j} \, ,
\end{equation}
where the walk operator
\begin{equation}
\label{eq:the_walk_op}
W^{(n)} \defeq S^{(n)} C^{(n)} \, ,
\end{equation}
is the composition of (i) a possibly position-dependent total coin operator
\begin{equation}
C^{(n)} \defeq \sum_{k=0}^{N-1} \ket{k} \! \! \bra{k} \otimes {C}_{k} \, ,
\end{equation}
and of (ii) a coin-dependent shift operator,
\begin{equation}
\begin{split}
S^{(n)} &\defeq \sum_{k=0}^{N-1} \Big( \ket{k-1 \ \text{mod} \ N} \! \! \bra{k} \otimes \ket{0} \! \! \bra{0}   \\
& \ \ \ \ \ \ \ \ \   + \ket{k+1 \ \text{mod} \ N} \! \! \bra{k} \otimes \ket{1} \! \! \bra{1} \Big) \, .
\end{split}
\end{equation}
Each $C_k$ is a $2 \times 2$ matrix acting on the coin Hilbert space, which we also call ``coin operator''\footnote{Notice that both $C^{(n)}$ and $C_k$ are called ``coin operator'', but the context should make it clear whether we refer to the former or the latter whenever we speak of a coin operator. Sometimes, we will specify ``total coin operator'' for $C^{(n)}$, as above.}.

\subsection{Encoding the position in base $2$}
\label{subsec:encoding}

We encode the position of the walker in base $2$ with $n\equiv \log_2(N)$ qubits, that we call \emph{position qubits}.
Let $k_2$ be the writing of $k$ in base $2$ using $n$ digits.
We will use the notations\footnote{It should be clear from the context whether some position state $\ket K$ means that $K$ is the writing of $k$ in base 10 or in base 2.}
%
%
\begin{equation}
\ket k \equiv \ket{k_2} \equiv \ket{b_{n-1} ... b_0} \, ,
\end{equation}
where $b_p = 0$ or $1$ with $p=0,...,n-1$, such that $k = \sum_{p=0}^{n-1} b_p \times 2^p$.

In the circuit, there are $n$ \emph{position wires}, each of which carries one position qubit, and these wires are, as the associated qubits, numbered from $0$ to $n-1$, starting from the bottom wire.
With $n$ qubits, we can encode $N = 2^n$ position states $\ket{k_2}$.
The coin state is encoded by an additional qubit.
In the circuit, there is thus an additional wire that we call \emph{coin wire}, which carries what we call the \emph{coin qubit}.
The coin wire is located at the bottom of the circuit.

We can thus write $S^{(n)}$ as
\begin{equation}
\begin{split}
{S}^{(n)} &\equiv \sum_{k=0}^{2^n-1} \Big( \ket{(k-1 \ \text{mod} \ N)_2} \! \! \bra{k_2} \otimes \ket{0} \! \! \bra{0}   \\
& \ \ \ \ \ \ \ \ \   + \ket{(k+1 \ \text{mod} \ N)_2} \! \! \bra{k_2} \otimes \ket{1} \! \! \bra{1} \Big) \, ,
\end{split}
\end{equation}
and ${C}^{(n)}$ as
\begin{equation}
\label{eq:non-uniform_coin_operator}
{C}^{(n)} \equiv \sum_{k=0}^{2^n-1} \ket{k_2} \! \! \bra{k_2} \otimes \tilde{C}_{k_2}  \, ,
\end{equation}
where, simply,
\begin{equation}
\tilde{C}_{k_2} \defeq C_k \, .
\end{equation}

\section{Quantum circuits for the shift operator}
\label{sec:shift}

In this section, we first (i) review (Sec.\ \ref{subsec:QFT_scheme}) a certain known scheme to implement with a quantum circuit the coin-dependent shift operator $S^{(n)}$, based on the quantum Fourier transform (QFT), and which we call QFT scheme \cite{Shakeel2020a}, and then (ii) comment (Sec.\ \ref{subsec:different_schemes}) on other existing schemes, and conclude on the advantage brought by the use of the QFT scheme.

\subsection{Quantum-Fourier-transform scheme}
\label{subsec:QFT_scheme}

In this section, we briefly present the scheme introduced in Ref.\ \cite{Shakeel2020a} (referred to as ``the reference'' in this subsection)
in order to implement with a quantum circuit the coin-dependent shift operator $S^{(n)}$,
where the key point is the use of the QFT.
(This idea was already introduced in Ref.\ \cite{QLMAplus2016} for the CQW.)
First, the authors combine the Hilbert spaces corresponding to (i) the coin degree of freedom, which is referred to as the ``velocity'',
$v=0,1$ (meaning direction of displacement), and to (ii) the position $k$, so as to obtain
states of the form $\ket{v,k} \defeq \ket{v} \otimes \ket{k} \equiv \ket{c} \otimes \ket{k}$, which span the whole Hilbert space.
They use this order in the tensor product because they work with the convention (which we also use in our paper) that what is on the right is embedded into what is on the left, i.e., here, the position space is embedded into the coin space, and this embedding is convenient here\footnote{In the rest of our paper, see already Sec.\ \ref{sec:1} above, we rather write $\ket{k} \otimes \ket{c}$ because if the coin operator is position dependent (which is the point of our work), then it is convenient to embed the coin Hilbert spaces into the position Hilbert space.}, see why just below in Eq.\ \eqref{eq:SNmatrix}.
Using this joint basis, $(\ket{v,k})_{v=0,1;k=0,...,N-1}$, one can write, in matrix notation, and with the aforementioned embedding,
\begin{equation}
S^{(n)}=\left[\begin{array}{cc}
I_{N} & 0\\
0 & J
\end{array}\right]\left[\begin{array}{cc}
S_{\text{R}} & 0\\
0 & S_{\text{R}}
\end{array}\right]\left[\begin{array}{cc}
I_{N} & 0\\
0 & J
\end{array}\right],\label{eq:SNmatrix}
\end{equation}
where all submatrices have dimensions $N\times N$, with (i) $I_N$ the identity matrix of size  $N\times N$, (ii) $S_{\text{R}}$
implementing the right shift operation $S_{\text{R}} \ket{k} \defeq \ket{k+1\ \text{mod}\ N}$ (in the reference, $S_{\text{R}}$ is denoted by $\mathbf{X}$),
and (iii) $J$ the exchange matrix with all its entries equal to $0$ except those
on the anti-diagonal, which are all $1$'s.
$S^{(n)}$ is denoted by $\sigma$ in the reference.
Each exchange matrix
is implemented using $n$ CNOT gates (see Fig.\ 1 in the reference).
Now, the authors suggest implementing the right shift operator by first performing a QFT, namely,
\begin{equation}
\mathcal{F}:\ket{k} \mapsto \frac{1}{\sqrt{N}} \sum_{q=0}^{N-1}e^{2\pi ikq/N}\ket{q} \, .
\end{equation}
The reason for using the QFT is that it can be regarded as a change of basis, so that in this new basis the right shift operator is diagonal. 
That is to say, one has
\begin{equation}
S_{\text{R}}=\mathcal{F}^{-1}\Omega\mathcal{F} \, , 
\end{equation}
where $\Omega$ is the diagonal matrix with entries $e^{2\pi iq/N},q=0,\dots N-1$.
For $N=2^n$, the quantum circuit for $\Omega$ can be obtained as a tensor product of single-qubit rotations of the form
\begin{equation}
R_{l}\defeq\begin{bmatrix}1 & 0\\
0 & e^{2\pi i/{2^{l}}} 
\end{bmatrix} \, ,
\end{equation}
with $l=1, ..., n-1$, see Eq.\ (12) of the reference.
In the reference, the quantum circuit implementing the quantum Fourier transform is given in Fig.\ 2, and the quantum circuit finally implementing $S^{(n)}$ (denoted by $\sigma$ in the reference) is given in Fig.\ 5 (we omit the application of the coin operator, denoted by $\hat{S}$ in the reference).

\subsection{Comments on other existing schemes}
\label{subsec:different_schemes}

On the one hand, one needs the circuit depth to be as small as possible,
as we are limited by the decoherence time of the qubits. On the other
hand, we also want to reduce the number of 2-qubit gates (and of course, more generally, of multi-qubit gates) because:
i) these gates have higher error rates, and ii) more importantly,
because if the qubits that are involved in a 2-qubit gate are not
directly linked in the chip, they must be moved to adjacent positions (usually
using SWAPs) which increases the number of gates and the total error
rate. 
More generally, we want to decompose any multi-qubit unitary operation into a minimum number of one-qubit and two-qubit gates. 

Many circuit implementations of the shift operator $S^{(n)}$ make use of the
increment and decrement (ID) scheme introduced in Ref.\ \cite{FOBH2005},
which is explained in Appendix \ref{app:Fuji}.
Following Ref.\ \cite{Shakeel2020a}, one can show that for the ID scheme one needs a circuit of size $n(2n^{2}-6n+7)/3$ and
depth $2(2n^{2}-8n+9)$.
Alternatively, $n-3$ extra ancillary qubits
can be used, so that the circuit size becomes $10n^{2}-50n+67$ and the depth $2(4n^{2}-20n+27)$.
As compared to these quantities, the QFT scheme (which we have described above in Sec.\ \ref{subsec:QFT_scheme}) makes use of a circuit of size $n^{2}+4n+1$ and depth
$2n+3$ for the case of $N=2^{n}$ lattice sites.
Thus, the use of
the QFT scheme represents a clear advantage. 

In Ref.\ \cite{GEZ2021}, a different scheme is introduced that
uses rotations around the basis states.
They conclude that the number
of gates needed by their scheme increases exponentially
faster than for the generalized ID scheme (the one which
includes ancillary qubits) -- although they also point out that the latter scheme anyways quickly demands a number of qubits that surpasses that available in near-term quantum machines.
Let us also mention that just modifying the digital mapping of the position space might be used to reduce the required gate count for implementing DQWs, as illustrated by some examples in Ref.\ \cite{SAHBplus2021}.

\section{Naive quantum circuit for position-dependent coin operator}
\label{sec:naive}

\begin{figure*}[t]
\hspace{-0.5cm}
    \includegraphics[width=8cm]{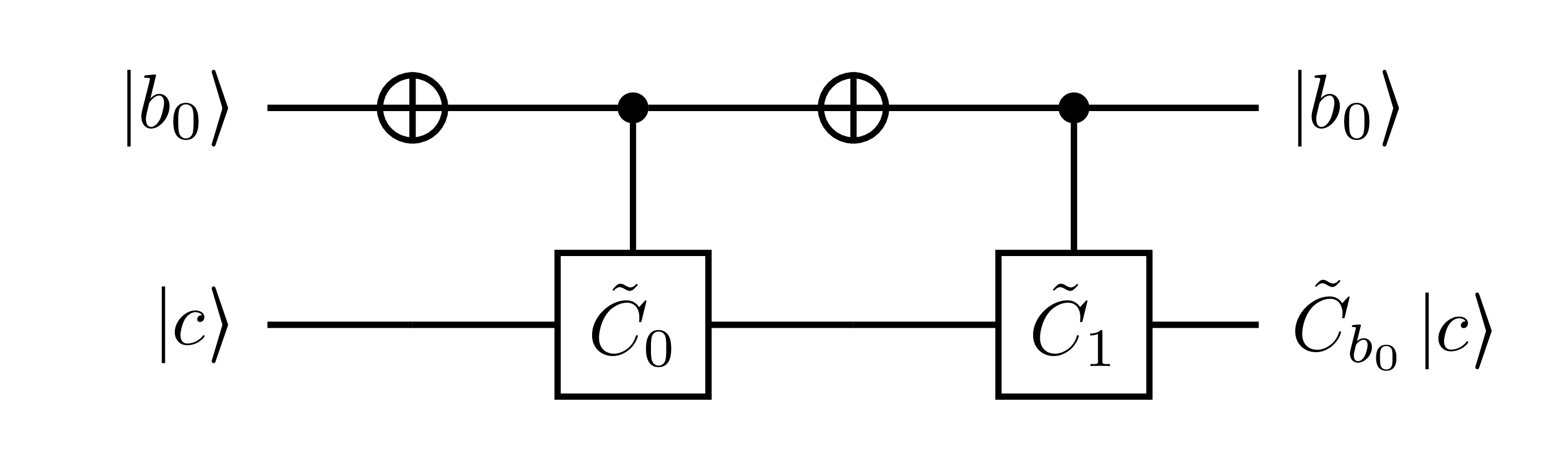} \ \ \ \ \
    \includegraphics[width=9.4cm]{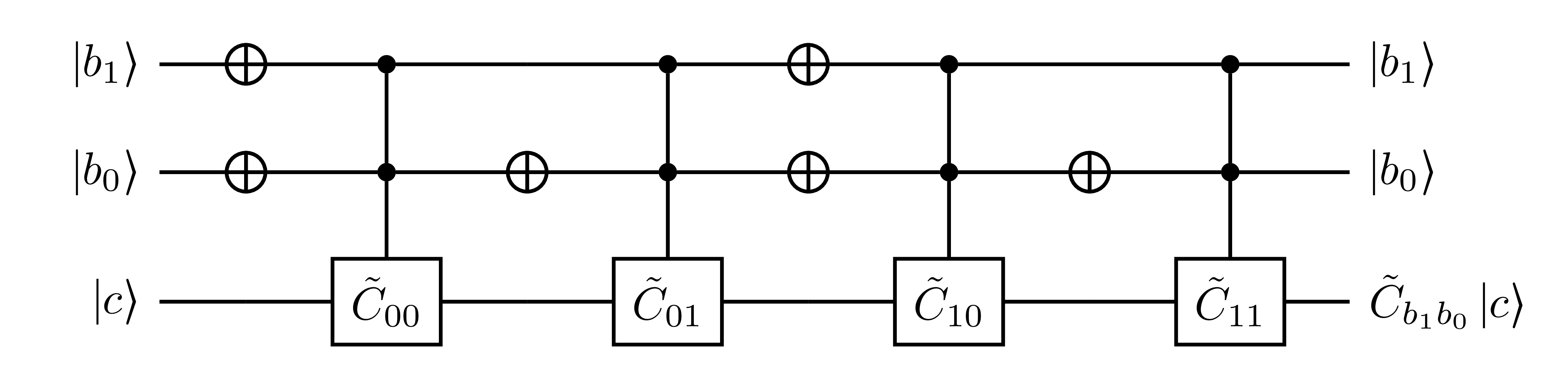}
    \includegraphics[width=18cm]{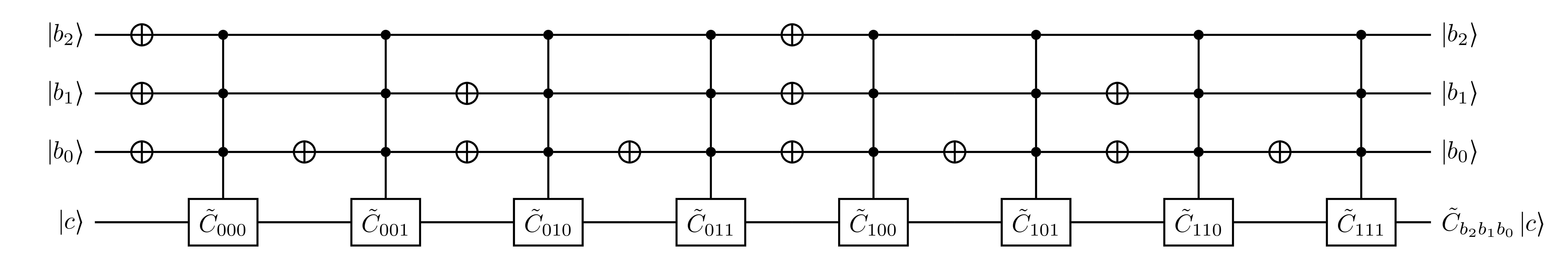}
    \includegraphics[width=18cm]{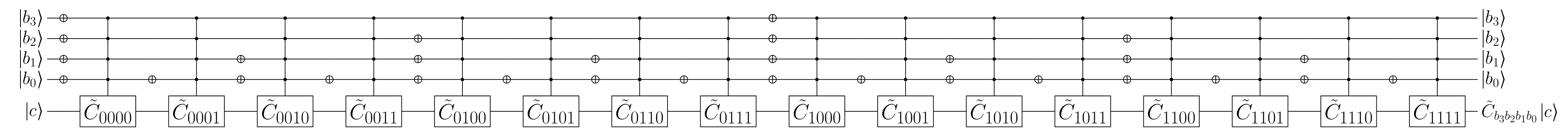}
    \caption{Circuits implementing the position-dependent coin operator $U^{(n)}={C}^{(n)}$ for $n=1$ (top left figure), $n=2$ (top right figure), $n=3$ (middle figure), and $n=4$ (bottom figure).}
    \label{fig:position-dependent_coin_operator}
\end{figure*}

We want to implement ${C}^{(n)}$, defined in Eq.\ \eqref{eq:non-uniform_coin_operator}.
A first, naive idea that comes to one's mind is to place all the coin operators $\tilde{C}_{k_2}$ sequentially, i.e., one after the other, along the coin wire.
This already implies that the depth of the circuit will be exponential in the number of position wires (this will become clearer further down).
The coin operator $\tilde{C}_{k_2}$ applied to the coin qubit must be that corresponding to the position on which the walker is initially, i.e., before applying our circuit. 
Let us then control each coin operator on all position wires, which means that we apply sequentially the $2^n$ gates $G_n(C_k)$, where $G_n(M)$ is the application of the gate $M$ on the coin qubit, controlling it by the $n$ position qubits of the circuit; In Eqs.\ \eqref{eqs:Gs} below we define $G_n(M)$ explicitly, mathematically, by induction. 
Our task is now that the state carried by all the position wires at the level of a coin operator $\tilde{C}_{k_2}$ be $\ket{1...1}$ ($n$ times $1$) if and only if (iff) the initial position state $\ket{b_{n-1}...b_0}$ is $\ket{k_2}$, so that $\tilde{C}_{k_2}$ is ``activated'', i.e., applied to the coin qubit, only in that case.
The circuit $U^{(n)}$ shown in Fig.\ \ref{fig:position-dependent_coin_operator} for $n=1$ to $4$, does the job, i.e., 
\begin{equation}
\label{eq:The_equality}
U^{(n)} = {C}^{(n)} \, .
\end{equation}
We will first give an informal proof of Eq.\ \eqref{eq:The_equality}, and then a formal one.
The informal proof is also a way to explain how the circuit $U^{(n)}$ is built.
In these circuits, the circles $\oplus$ are, as usual, NOT gates, i.e., gates
\begin{equation}
X\defeq \begin{bmatrix}
0 & 1 \\ 1 & 0 \end{bmatrix} \, .
\end{equation}
As announced earlier and as it can be seen on Fig.\ \ref{fig:position-dependent_coin_operator}, the construction outlined above implies a depth of the circuit that is exponential in the number of position wires.

\subsection{Informal proof / construction of $U^{(n)}$}
\label{subsec:construction}

The informal proof is the following. 
We do it for $n=3$, see Fig.\ \ref{fig:position-dependent_coin_operator}, middle figure.
Let us start from the beginning of the circuit, i.e., on the left.
The coin operator $\tilde{C}_{000}$ is activated iff the running position state is $\ket{111}$, i.e., iff the initial position state is $\ket{000}$, which is what we want.
Before proceeding with the next coin operators, let us check that at the end of the circuit the initial position state is unchanged: after $\tilde{C}_{000}$ there are, on each position wire, an odd number of $X$ gates, which reduces to one $X$ gate per wire, which cancels out with the ``tower'' of initial $X$ gates (``initial'' here means positioned before $\tilde{C}_{000}$), so that the position state is left unchanged at the end of the circuit. This check is of course valid for any initial position state.
%

Let us now proceed.
The coin operator $\tilde{C}_{001}$ is activated iff the running position state is $\ket{111}$, i.e., iff the state at $\tilde{C}_{000}$ is $\ket{110}$, i.e., iff the initial state is $\ket{001}$, which is what we want.
Let us reformulate the first step of the previous reasoning: after having tested whether $\tilde{C}_{000}$ is activated by the associated position state, we test if modifying this position state $k$ by $1$ activates the following coin operator, $\tilde{C}_{001}$; this modification requires, in base $2$, flipping the first position qubit, that is, applying an $X$ gate on the first position wire.
%

By continuing the above reasoning, one may  convince oneself  that the circuit does the job. In particular, one may convince oneself that (i) as we have explained above, if we start by $\ket{b_{n-1}...b_0}$, then as we advance in the circuit none of the coin operators will be activated unless it is $\tilde{C}_{b_{n-1}...b_0}$, and also that (ii) \emph{none of the remaining coin operators will be activated}.

The construction of the circuit can be summed up as follows: between one $n$-controlled coin operator $G_n(C_k)$ and the next one $G_n(C_{k+1})$, we apply ``a tower of'' $X$ gates,
\begin{equation}
T_n(k \ \text{mod} \ 2^{n-1}) = \underbrace{I_2 \otimes ... \otimes I_2}_{n-h} \otimes \underbrace{X \otimes ... \otimes X}_{h} \otimes I_2 \, ,
\end{equation}
whose height $h$ is determined by the following reasoning: we place an $X$ gate on the position wire $p$ (from $0$ to $n-1$ starting from the lowest wire) iff the bit $b_p$ of $k_2$ changes by going from $k_2$ to $(k+1)_2$.

\begin{figure*}[t]
\hspace{-0.5cm}
    \includegraphics[width=18cm]{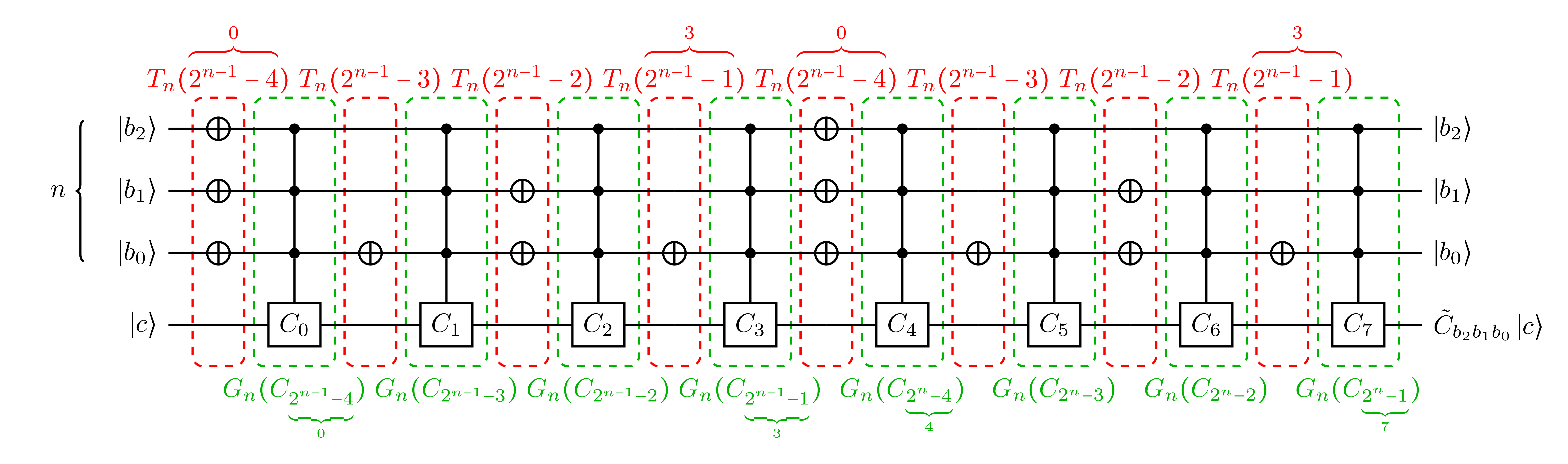}
    \caption{Circuit corresponding to Formula \eqref{eq:the_circuit} for $n=3$, implementing the position-dependent coin operator $U^{(n)}={C}^{(n)}$, see Eq.\ \eqref{eq:The_equality}.}
    \label{fig:the_formula}
\end{figure*}

\subsection{Explicit definition of $U^{(n)}$}

The circuit $U^{(n)}$ the construction of which is explained in Sec.\ \ref{subsec:construction} above, is defined by the following formula,
\begin{equation}
\label{eq:the_circuit}
U^{(n)} \defeq \sideset{}{^L}\prod_{k=0}^{2^n-1} G_n(C_k) T_n(k \ \text{mod} \ 2^{n-1}) \, ,
\end{equation}
where we define $G_n(C_k)$ and  $T_n(i)$ by induction,
\begin{subequations}
\begin{align}
T_1(0) &\defeq X \otimes I_2 \\
T_n(i) &\defeq M_i \otimes T_{n-1}(i \ \text{mod} \ 2^{n-2}) \, , \label{eq:towers}
\end{align}
\end{subequations}
with
\begin{equation}
M_i \defeq \left\{
\begin{array}{ll}
X & \text{if} \ i = 0 \\
I_2 & \text{otherwise}
\end{array}\right. \, ,
\end{equation}
and
\begin{subequations}
\label{eqs:Gs}
\begin{align}
G_1(C_k) &\defeq I_2 \oplus C_k = \begin{bmatrix}
I_2 & 0 \\ 0 & C_k
\end{bmatrix} \\
G_n(C_k) &\defeq I_{2^n} \oplus G_{n-1}(C_k) = \begin{bmatrix}
I_{2^n} & 0 \\ 0 & G_{n-1}(C_k) 
\end{bmatrix} \, . \label{eq:Gs}
\end{align}
\end{subequations}
The superscript ``$L$'' in the product of Eq.\ \eqref{eq:the_circuit} means that each term of the product is multiplied \emph{on its left} by the next one.
The formula of Eq.\ \eqref{eq:the_circuit} is illustrated in Fig.\ \ref{fig:the_formula} for $n=3$.
In Appendix \ref{app:proof}, we prove by induction on $n$ that $U^{(n)}$ defined by Eq.\ \eqref{eq:the_circuit} is equal to ${C}^{(n)}$.

\section{Linear-depth quantum circuit for position-dependent coin operator}
\label{sec:linear}

The depth of the naive circuit presented Sec.\ \ref{sec:naive} above is exponential in the number of position wires.
Since the coherence time of the qubits is a high limiting factor in actual implementations of quantum circuits, it is known that whenever possible it is preferable to reduce the depth of a circuit at the cost of adding ancillary qubits\footnote{That being said, such a way to go implies that we need to handle many more entangled qubits; ultimately, this is only completely guaranteed if error correction is available.}.
In this section, we describe a quantum circuit $U^{(n)}_{\text{lin.}}$ implementing the position-dependent coin operator ${C}^{(n)}$, which has a linear depth in the number $n$ of position wires, at the cost of introducing an exponential number of ancillary wires and, correspondingly, an exponential number of gates.

\subsection{Adding the ancillary wires}

\begin{figure}[t]
\hspace{-0.7cm}
	\includegraphics[width=8cm]{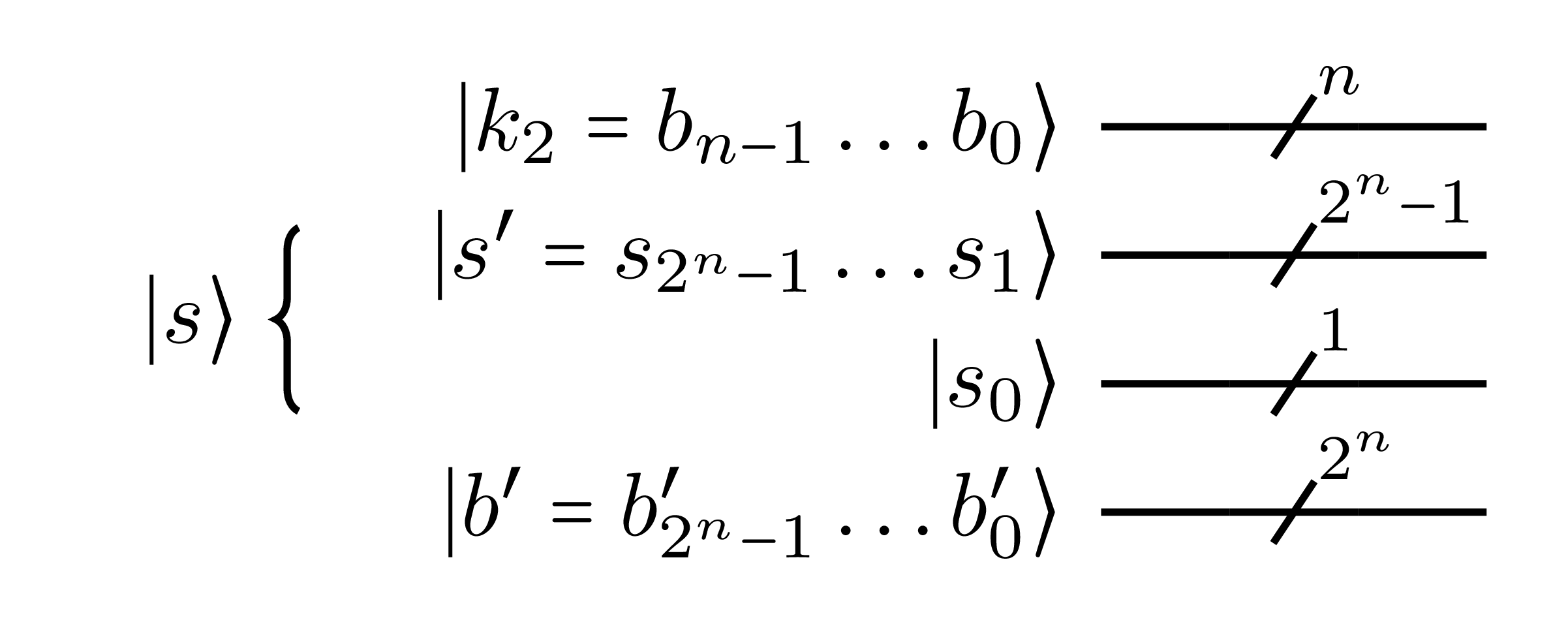}
	\caption{Registers necessary for the implementation of the linear-depth quantum circuit for the position-dependent coin operator. From top to bottom, there is (i) the positions register $\mathcal{H}_{\text{pos.}}$, (ii) the ancillary-coins register $\mathcal{H}'_{\text{coins}}$, (iii) the principal coin register $\mathcal{H}_0$, and (iv) the ancillary-positions register $\mathcal{H}'_{\text{pos.}}$. \label{fig:wires}}
\end{figure}

To the previous circuit with the $n$ position wires and the coin wire, we add the following ancillary wires: first, (i) $2^n-1$ \emph{ancillary coin wires} in between the position wires and the original coin wire which we call \emph{principal coin wire}, labelled with increasing index from bottom to top, from $1$ (the coin wire is labelled by $0$) to $2^n-1$, and (ii) $2^n$ \emph{ancillary position wires} below the principal coin wire, labelled with increasing index from bottom to top, from $0$ to $2^n-1$.
The collection of all position wires (meaning, more precisely, all possible superpositions of position states) is called positions register, and is simply the Hilbert space $\mathcal{H}_{\text{pos.}}$, that we have already defined in Sec.\ \ref{subsec:Hilbert}.
The collection of all ancillary position wires is called ancillary-positions register, and is a Hilbert space $\mathcal{H}_{\text{pos.}}'$.
The collection of (i) all ancillary coin wires is called ancillary-coins register, which is a Hilbert space $\mathcal{H}_{\text{coins}}'$, and the collection of (ii) all coin wires (both the principal coin wire and the ancillary-coins wires) is called coin register, which is the Hilbert space $\mathcal{H}_{\text{coins}} \defeq \mathcal{H}_{\text{coins}}' \otimes \mathcal{H}_{0}$,

We have seen in Sec.\ \ref{subsec:encoding} above that a position state is denoted by $\ket{k_2} = |b_{n-1}...b_0\rangle$, with $b_p=0$ or $1$, $p=0,...,n-1$, and that the coin state is denoted by $\ket{c}$, which can be in any superposition of the up and down states.
A basis state of the ancillary-positions register is denoted by $\ket{b'}$, and is made up of a string of bits $b'_m=0$ or $1$, $m=0,...,2^n-1$, that is,
\begin{equation}
\ket{b'} \defeq \ket{b'_{2^{n}-1}...b'_0} \, .
\end{equation}
A basis state of the coin register is denoted by $\ket{s}$, and is made up of a string of bits $s_m=0$ or $1$, $m=0,...,2^n-1$ -- where $m=0$ labels a basis state of the principal coin qubit --, that is,
\begin{equation}
\ket{s} \defeq \ket{s_{2^{n}-1}...s_0} \, .
\end{equation}
We also use the notation
\begin{equation}
\ket{s'} \defeq \ket{s_{2^{n}-1}...s_1} \, .
\end{equation}
Notice the use of a number of ancillae which is exponential in the number $n$ of position wires.
Figure \ref{fig:wires} illustrates the different registers of the circuit.

\subsection{Main idea and definition of $Q_0$}

\begin{figure}[t]
\hspace{-0.7cm}
	\includegraphics[width=6cm]{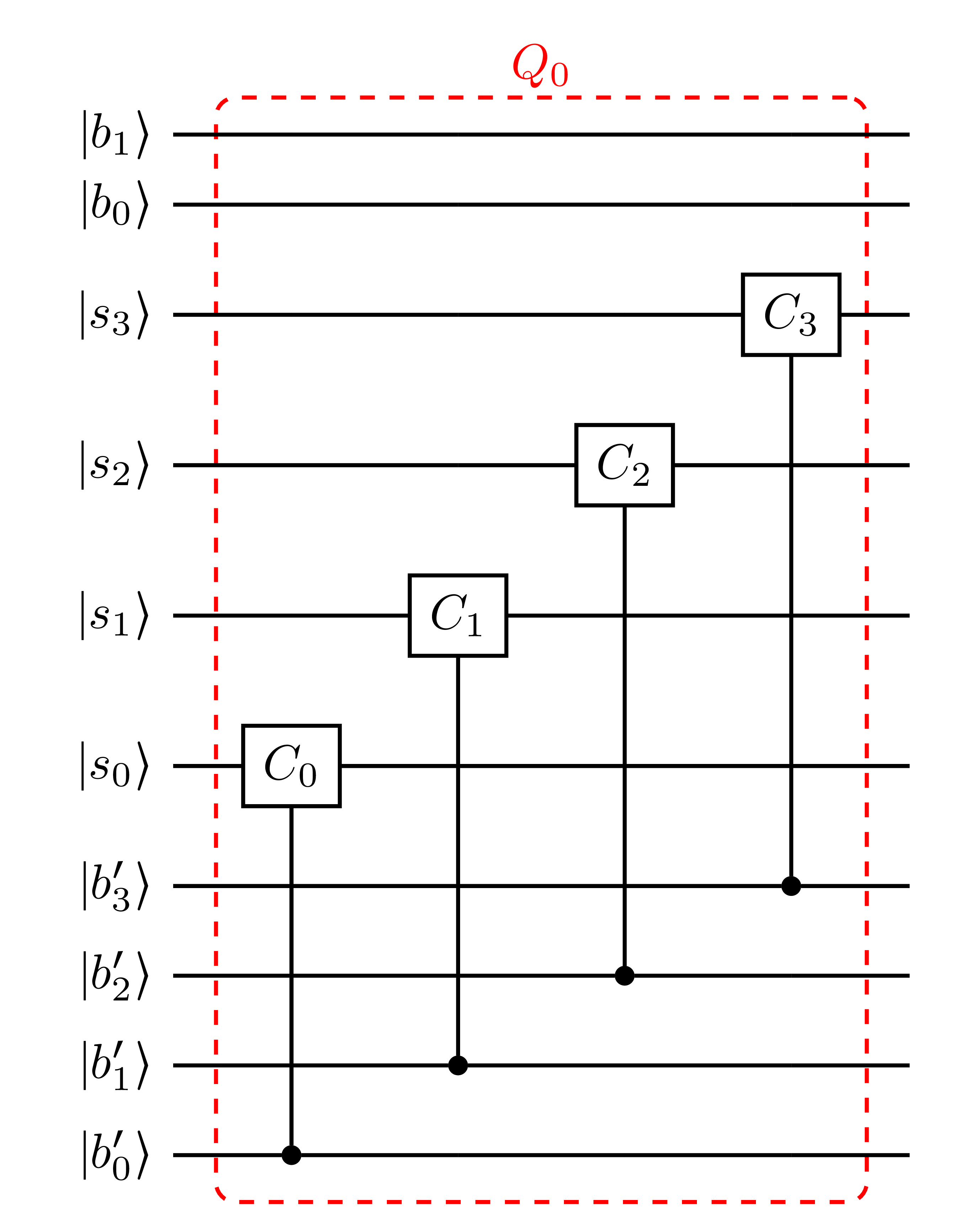}
	\caption{Main operation, called $Q_0$, of the linear-depth quantum circuit implementing the position-dependent coin operator defined in Eq.\ \eqref{eq:non-uniform_coin_operator}, for $n=2$. We see that all controlled-$C_k$ operations can by applied simultaneously. The explicit definition of $Q_0$ is given in Eq.\ \eqref{eq:Q0}. \label{fig:Q0}}
\end{figure}

The main idea of the circuit we want to build is to apply all the coin operators $C_k$ in parallel, instead of doing it sequentially as in the naive circuit of Sec.\ \ref{sec:naive}.
This is why we have added $2^n-1$ ancillary coin wires, i.e., on each of those wires we are going to apply a $C_k$.
Now, the application of $C_k$ should be done only if the position state is the corresponding one, that is, $\ket k$.
This is why we have added $2^n$ ancillary position states: each application of $C_k$ on the coin wire number $k$, is controlled on the ancillary position wire number $k$.
All these 1-controlled coin operations can thus be applied in parallel, and they form the central operation of our circuit, which we call $Q_0$, depicted in Fig.\ \ref{fig:Q0} for $n=2$.

The explicit definition of $Q_0$ is the following:
\begin{equation}
\label{eq:Q0}
Q_0 \defeq I_{2^n} \otimes \left( \sideset{}{^L}\bigotimes_{k=0}^{2^n-1} K_{b'_k,s_k}(C_k) \right) \, ,
\end{equation}
where $K_{b'_k,s_k}(M)$ is the controlled-$M$ operation which controls on the qubit $\ket{b'_k}$ and applies $M$ to the qubit $\ket{s_k}$, and where we recall that the superscript ``$L$'' on the tensor product means that each term of the product is multiplied on its left by the next term\footnote{We give this precision to keep coherence with how we handle the Kronecker product in this paper, but actually here in Eq.\ \eqref{eq:Q0} it is not possible to keep the strict order that what is at the bottom of the circuit goes on the right in the tensor product and what is a the top goes on the left because the controlled-$C_k$ operations entangle qubits which are separated by several wires on the circuit.}.
The big tensor product of Eq.\ \eqref{eq:Q0} indicates that all the terms of the product can be applied in parallel, i.e., at once, which is the main point of the circuit we want to build.
$Q_0$ is trivially unitary since all the factors in the tensor products are unitary.
In Eq.\ \eqref{eq:Q0}, we have omitted many identity tensor factors in order to lighten the writing, except from the $I_{2^n}$ which acts on the position qubits, left untouched during the application of $Q_0$.

\subsection{Initialization of the ancillae via $Q_1$ and $Q_2$, and final scheme $U^{(n)}_{\mathrm{lin.}}$ implementing ${C}^{(n)}$}

\begin{figure}[t]
\hspace{-0.5cm}
	\includegraphics[width=9cm]{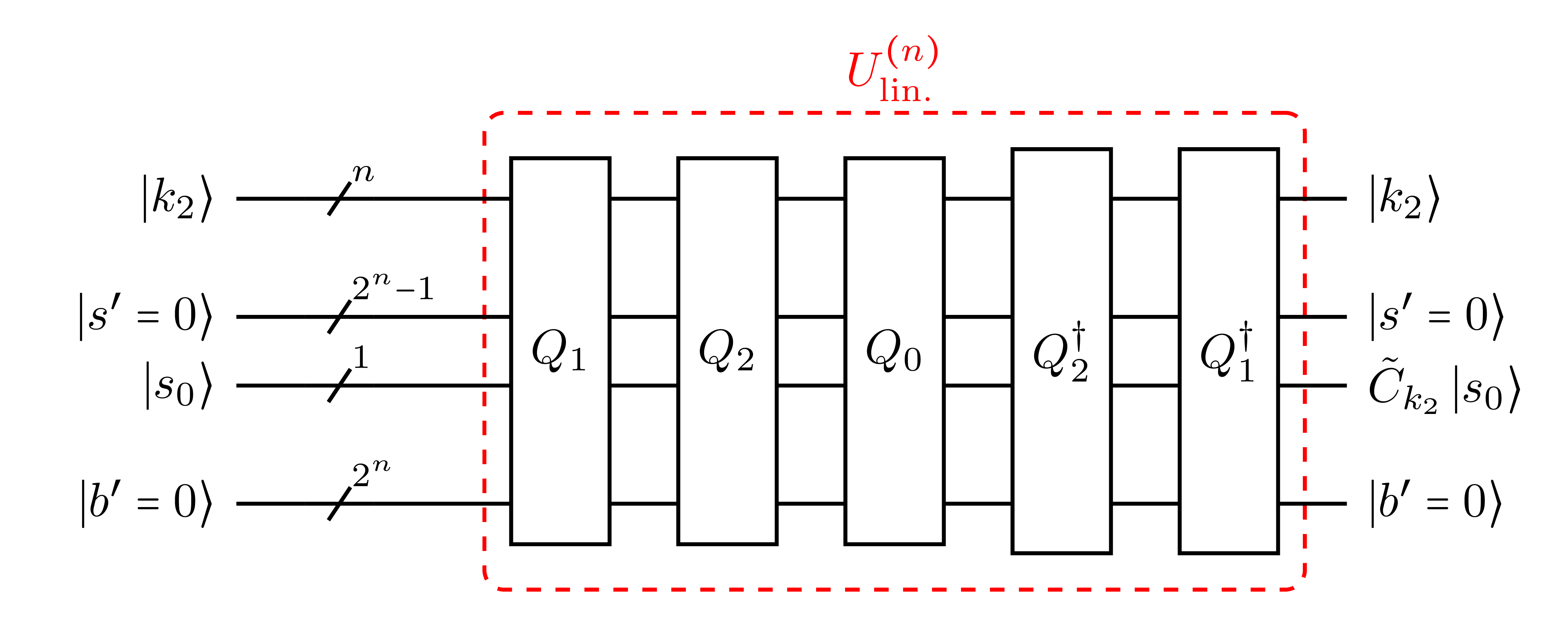}
	\caption{Linear-depth quantum circuit $U^{(n)}_{\text{lin.}}$, defined in Eq.\ \eqref{eq:U_linear}, implementing the position-dependent coin operator ${C}^{(n)}$, defined in Eq.\ \eqref{eq:non-uniform_coin_operator}, that is, we have that $U^{(n)}_{\text{lin.}} = {C}^{(n)} \otimes I_{2^{(2^n-1)}} \otimes I_{2^{(2^n)}}$, where ${C}^{(n)}$ acts on the tensor product $\mathcal{H}$ of the  position space and the principal-coin space, $I_{2^{(2^n-1)}} $ acts on the ancillary coins  $\mathcal{H}'_{\text{coins}}$, and $I_{2^{(2^n)}}$ on the ancillary positions $\mathcal{H}'_{\text{pos.}}$. \label{fig:U_linear}}
\end{figure}

In order to apply $Q_0$ meaningfully, two things must be done.
First, before applying $Q_0$, that is, all coin operations in parallel, one must have loaded the appropriate coin state on each ancillary coin wire, an operation which we call $Q_2$, and after applying $Q_0$ one must load back the new information contained in the ancillary coin wires onto the principal coin wire, which is done by $Q_2^{\dag}$, an operation which also restores on the ancillary coin wires the states that were there before applying $Q_2$.
This means that instead of applying directly $Q_0$ we must apply $Q_2^{\dag}Q_0 Q_2$.

A second thing that must be done before applying $Q_2^{\dag}Q_0 Q_2$, is to make sure that each ancillary position qubit $b'_{k}$ is $1$ if and only if the position qubits encode $k$, i.e., correspond to $\ket{k_2}$, an operation which we call $Q_1$, and after applying $Q_2^{\dag}Q_0 Q_2$ one must restore the initial states of the ancillary position wires, which is done by $Q_1^{\dag}$.
The final scheme that we must apply thus reads
\begin{equation}
\label{eq:U_linear}
U^{(n)}_{\text{lin.}} \defeq   Q_1^{\dag}  Q_2^{\dag} Q_0 Q_2 Q_1 \, ,  
\end{equation}
depicted in Fig.\ \ref{fig:U_linear}.

We define $Q_1$ and $Q_2$ in Appendices \ref{app:Q1} and \ref{app:Q2}, respectively.
In Appendix \ref{app:proof_linear} we prove that the circuit $U^{(n)}_{\text{lin.}}$ of Eq.\ \eqref{eq:U_linear} indeed coincides with the position-dependent coin operator ${C}^{(n)}$ if both ancillary registers $\mathcal{H}'_{\text{coins}}$ and $\mathcal{H}_{\text{pos.}}'$ are initialized with the joint state $\ket{s'=0} \ket{b'=0}$: that is to say, we prove that
\begin{equation}
\label{eq:the_thing}
U^{(n)}_{\text{lin.}} \ket{S} = \left( {C}^{(n)} \otimes I_{2^{(2^n-1)}} \otimes I_{2^{(2^n)}}  \right) \ket{S} \, ,
\end{equation}
where $I_{2^{(2^n-1)}}$ acts on the ancillary coins $\mathcal{H}'_{\text{coins}}$, $I_{2^{(2^n)}}$ acts on the ancillary positions $\mathcal{H}'_{\text{pos.}}$, and where the most general form for $\ket S$ is to be arbitrary on $\mathcal{H}$ and equal to $\ket{s'=0} \ket{b'=0}$ on $\mathcal{H}'_{\text{coins}} \otimes \mathcal{H}_{\text{pos.}}'$, that is,
\begin{equation}
\label{eq:special}
    \ket S \defeq \left( \sum_{k=0}^{2^n-1} \sum_{s_0=0,1} \alpha_{k,s_0} \ket{k_2}\ket{s_0} \right)  \ket{s'=0} \ket{b'=0} \, ,
\end{equation}
with the $\alpha_{k,s_0}$s being complex numbers such that $\sum_{k=0}^{2^n-1} \sum_{s_0=0,1} |\alpha_{k,s_0}|^2 = 1$.

Notice that in the diagrammatic representation of our circuit, see Fig.\ \ref{fig:U_linear}, we have positioned, from top to bottom, the position states, the ancillary coin states, the principal coin state, and the ancillary position states, whereas in the equations, see, e.g., Eq.\ \eqref{eq:special}, we have used the order $\ket{k_2}\ket{s_0}  \ket{s'=0} \ket{b'=0}$; we have done so on the diagrams because it makes the circuits more understandable, and we have done so on the equations because it is clearer to write all ancillary states together and after the non-ancillary ones.

\subsection{Exact depth of the circuit}

\begin{figure}[t]
\hspace{-0.0cm}
	\includegraphics[width=5cm]{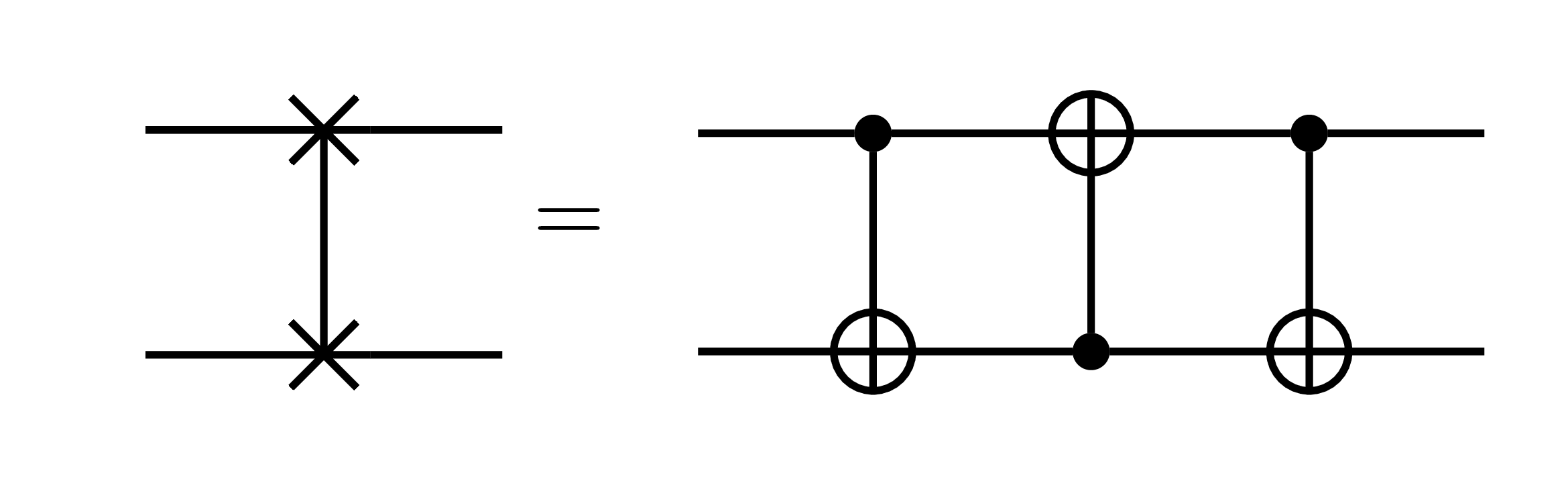}
	\caption{One possible expression of the SWAP operation in terms of CNOT gates.  \label{fig:swaps}}
\end{figure}

Let $d$ be the function that returns the depth of a circuit.
The depth of $U^{(n)}_{\text{lin.}}$ is:
\begin{subequations}
\begin{align}
d(U^{(n)}_{\text{lin.}}) &= d(Q_1) + d(Q_2) + d(Q_0) + d(Q_2^{\dag}) + d(Q_1^{\dag}) \\
&= 2 d(Q_1) + 2 d(Q_2) + d(Q_0) \, .
\label{eq:theeq}
\end{align}
\end{subequations}
We count $3$ for the depth of the SWAP operation, since it is given by $3$ consecutive CNOT gates, as shown in Fig. \ref{fig:swaps}.

Now, the depth of $Q_1$ is the following,
\begin{subequations}
\begin{align}
d(Q_1) &= d(Q_{10}) + d(Q_{11}) + d(Q_{10}^{\dag}) \\
&= 2 d(Q_{10}) + d(Q_{11}) \\
&= 2 \underbrace{(n-1)}_{\text{copies}} + \, 3 \underbrace{n}_{\text{SWAPs}} + \underbrace{1}_{\text{NOT gate} } \\
&= 5n - 1 \, .
\end{align}
\end{subequations}
But, when $Q_1$ contains the copies operation $Q_{10}$, that is, for $n\geq 2$, the first NOT gate (i.e., $X$ gate)  of $Q_{11}$, which is applied on $\ket{b'_0}$, can be applied in parallel with the copies operation, so that, actually,
\begin{equation}
\label{eq:1_}
d(Q_1)  = 5n - 2 + \delta_{1,n} \, ,
\end{equation}
where $\delta_{1,n}$ is the Kronecker symbol.
The depth of $Q_2$ is the following,
\begin{subequations}
\label{eqs:2_}
\begin{align}
d(Q_2) &= 2 \underbrace{(n-1)}_{\text{CNOTs}} + \, 3 \underbrace{n}_{\text{SWAPs}} \\
&= 5n - 2 \, .
\end{align}
\end{subequations}
Finally, the depth of $Q_0$ is
\begin{equation}
\label{eq:3_}
d(Q_0) = 1 \, .
\end{equation}

Inserting then Eqs.\ \eqref{eq:1_}, \eqref{eqs:2_} and \eqref{eq:3_} into Eq.\ \eqref{eq:theeq} yields
\begin{subequations}
\begin{align}
d(U_{\text{lin.}}^{(n)}) &= 2(5n -2 + \delta_{1,n}) + 2(5n - 2) +1 \\
&= 20n  + 2 \delta_{1,n} - 7 \\
&= O(n) \, .
\end{align}
\end{subequations}
Let us finally point out that, in the depth calculations, we have not taken into account the fact that certain final operations of one block, say, $Q_1$, may be done in parallel with the first operations of the next block, say, $Q_2$. 
Therefore, what we have computed is an upper bound for the depth.
That being said, the asymptotic complexity is not impacted by these discrepancies.

\section{Efficient quantum circuit for smooth position-dependent coin operator}
\label{sec:efficient}

In this section we consider a coin operator with smooth but otherwise arbitrary position dependence, and show that this smoothness condition enables its implementation with an efficient quantum circuit composed of only one-qubit and two-qubit gates.
The precise smoothness condition will be given further down in Sec.\ \ref{subsec:approx} and Appendix \ref{subapp:efficiency}.
That such a quantum circuit is efficient means that its size and depth scale as $O(\text{poly}(n,1/ \epsilon))$, where as before $n$ is the number of position qubits, and $\epsilon$ is the error between the implemented unitary and the targeted unitary in terms of the spectral norm.
Such coin operators with smooth position dependence are particularly relevant for both (i) DQWs coupled to gauge fields \cite{DDMEF12a, AD16a, AD16b, MMAMP18, CGWW18, AMBD16, DMD13b, DMD14, AD17, AF17}, and (ii) certain DQW spatial-search schemes \cite{ZD2021}.

\subsection{Introduction: decomposition of an arbitrary position-dependent coin operator}

As above, the position-dependent coin operator is given by Eq.\ \eqref{eq:non-uniform_coin_operator}, and it is known that one can parametrize the generic position-dependent $2 \times 2$ unitary matrix $\tilde{C}_{k_2}$ with the Euler angles and a global phase as
{\small
\begin{subequations}
\label{eq:arbitrary_coin_op}
\begin{align}
\tilde{C}_{k_2} &\equiv e^{\mathrm{i}\alpha(k_2)} 
\begin{bmatrix}
e^{\mathrm{i}\xi(k_2)} \cos \theta(k_2) & e^{\mathrm{i}\zeta(k_2)} \sin \theta(k_2) \\
-e^{-\mathrm{i}\zeta(k_2)} \sin \theta(k_2) & e^{-\mathrm{i}\xi(k_2)} \cos \theta(k_2) 
\end{bmatrix} \\
&= e^{\mathrm{i}\alpha(k_2)} e^{\mathrm{i}(\xi(k_2)+\zeta(k_2))\sigma^3} e^{\mathrm{i}\theta(k_2)\sigma^2} e^{\mathrm{i}(\xi(k_2) - \zeta(k_2)) \sigma^3} \\
&= e^{\mathrm{i}F_0(k_2)} e^{\mathrm{i}F_1(k_2)\sigma^3} e^{\mathrm{i}F_2(k_2)\sigma^2} e^{\mathrm{i}F_3(k_2) \sigma^3} \, ,
\label{eq:Ck2}
\end{align}
\end{subequations}
}
where $\sigma^2 \equiv Y$ is the second Pauli matrix, $\sigma^3 \equiv Z$ is the third Pauli matrix, and $F_0 \defeq \alpha$, $F_1 \defeq \xi + \zeta$, $F_2 \defeq \theta$, and $F_3 \defeq \xi -\zeta$.

For technical reasons, we assume that we have first encoded $k_2$, $k=0,...,2^n-1$, into a real value belonging to $[0,1]$, via a dyadic expansion, that is,
\begin{equation}
x_k \defeq g(k_2) \defeq  \sum_{p=0}^{n-1} b_p / 2^{p+1} \, ,
\end{equation}
where we recall that $\ket{k_2} \equiv \ket{b_{n-1}...b_0}$, and we then introduce differentiable functions $f_i$, $i=0,1,2,3$, defined on $[0,1]$, which interpolate the values $F_i(k_2)$, such that
\begin{equation}
F_i(k_2) \equiv f_i(g(k_2)) \equiv f_i(x_k) \, .
\end{equation}

We can hence write $\tilde{C}_{k_2}$ in Eq.\ \eqref{eq:Ck2} as
$ \tilde{C}_{k_2} = e^{\mathrm{i}f_0(x_k)} e^{\mathrm{i}f_1(x_k)\sigma^3} e^{\mathrm{i}f_2(x_k)\sigma^2} e^{\mathrm{i}f_3(x_k) \sigma^3}$, 
and the total coin operator $C^{(n)}$ as
\begin{equation}
\label{eq:Cn}
C^{(n)} = e^{\mathrm{i}\hat{f}_0\otimes I_2} e^{\mathrm{i}\hat{f}_1 \otimes \sigma^3} e^{\mathrm{i}\hat f_2 \otimes \sigma^2} e^{\mathrm{i} \hat f_3 \otimes \sigma^3}\, , 
\end{equation}
where the $\hat f_i$ are operators acting on the position Hilbert space which are diagonal in the position basis, such that
\begin{equation}
\hat f_i \ket{k_2} = f_i(x_k) \ket{k_2} \, .
\end{equation}

Let us now define
\begin{equation}
\label{eq:Cnfsigma}
C^{(n)}_{f,\sigma} \defeq e^{\mathrm{i} \hat f \otimes \sigma} \equiv 
\begin{bmatrix}
e^{\mathrm{i} f(x_0)\sigma }  & 0 & \cdots & 0 \\
0 & e^{\mathrm{i} f(x_2)\sigma }  & \cdots & 0 \\
\vdots & \vdots & \ddots & \vdots \\
0 & 0 & \cdots & e^{\mathrm{i} f(x_{2^n-1})\sigma } 
\end{bmatrix} \, ,
\end{equation}
where $\hat f$ is, as above, a diagonal operator in position space associated to the function $f$, and $\sigma$ is a $2 \times 2$ Hermitian matrix.
One can rewrite $C^{(n)}$ in Eq.\ \eqref{eq:Cn} in terms of several $C^{(n)}_{f,\sigma} $'s, as $C^{(n)} = C^{(n)}_{f_0,I_2} C^{(n)}_{f_1,\sigma^3} C^{(n)}_{f_2,\sigma^2} C^{(n)}_{f_3,\sigma^3} $.
We thus focus on constructing an efficient quantum circuit for  $C^{(n)}_{f,\sigma}$, which will require a smoothness condition on $f$.

\subsection{Approximate efficient quantum circuit for block-diagonal unitaries of the form $C^{(n)}_{f,\sigma}$}
\label{subsec:approx}

The case $\sigma = I_2$, and, more generally, the case of diagonal unitaries with smooth variations, has already been derived by J. Welch et al. \cite{WGMAG2014}.
Following their method, we generalize their result to block-diagonal unitaries of the form of $C^{(n)}_{f,\sigma} $ in Eq.\ \eqref{eq:Cnfsigma}, with a smooth-enough $f$, by using the development of $f$ into a so-called Walsh series.

The proof holds in Appendices  \ref{subapp:the_circuit} and \ref{subapp:efficiency}.
In Appendix  \ref{subapp:the_circuit} we first provide an exact circuit for $C^{(n)}_{f,\sigma}$ as the product  $C^{(n)}_{f,\sigma} = \prod_{j=0}^{2^n-1}U_j$, depending on the coefficients $a_j$ of the decomposition of $f$ as a Walsh series.
The depth of such a circuit is, as the previous product formula reminds, exponential in $n$.
Now, in Appendix \ref{subapp:efficiency} we explain that we can truncate the previous product at some $j=2^m$ such that the resulting size and depth of the approximating circuit $C^{(n)}_{f_m,\sigma} = \prod_{j=0}^{2^m-1}U_j$ scale as $O(1/\epsilon)$, where $\epsilon$ depends on $m$ and is, in terms of the spectral norm, the error committed when approximating  $C^{(n)}_{f,\sigma} $ by $C^{(n)}_{f_m,\sigma}$, and where we have introduced the truncation $f_m \defeq \sum_{j=0}^{2^m-1} a_j w_{j} $ of the development of $f$ into a Walsh series.
This implementation is efficient according to the standard definition, \cite{Childs2004thesis} since the scaling of the circuit, $O(1/\epsilon)$, satisfies the condition $1/\epsilon = O(\text{poly}(n,1/\epsilon))$.
Now, it is trivial to realize from the above explanations that the condition for the approximating circuit to be efficient is $m \ll n$, which implies (see Appendix \ref{subapp:efficiency}) a smoothness condition on $f$, namely,
\begin{equation}
\label{eq:smoothness_cond}
\text{sup}_{x\in [0,1]} f'(x) \ll \epsilon 2^n \, .
\end{equation}
Notice that $\epsilon$ depends on $m$ and so is independent of $n$ only up to a certain extent, namely, we do need the condition $m \ll n$ to be satisfied.


A final comment is in order here.
If we do not truncate the Walsh series, the full Walsh decomposition of the coin operator, namely, $C^{(n)}_{f,\sigma} = \prod_{j=0}^{2^n-1}U_j$, provides a quantum-circuit implementation of the coin operator the depth and size of which scale as $O(2^n)$. 
A priori, this implementation actually outperforms polynomially the naive implementation that we have presented earlier in Sec.\ \ref{sec:naive}, since the depth and size of the latter scale as $O(\text{poly}(n) \times 2^n)$, where the $\text{poly}(n)$ comes from the fact that we have to decompose the multiply-controlled gates into one- and two-qubit gates.
That being said, the naive circuit is still interesting because it is more intuitive than the Walsh-decomposition circuit, since it makes no use of the Walsh series, which is a rather technical tool.

\subsection{Exact efficient quantum circuit for a linear position dependence}

In the particular case of a function $f$ that is linear in $x_k$, the associated coin operator $C^{(n)}_{f,\sigma} $ can be implemented exactly using only $n$ two-qubit quantum gates, see Appendix \ref{subapp:linear_pos_dep}.
This can for example be used for the quantum simulation of $d$-dimensional DQWs, $d \geq 2$, with a constant magnetic field\footnote{The case of a constant electric field (with $d\geq 1$) can already be treated with Ref.\ \cite{WGMAG2014} in the spatial gauge since we only have a position-dependent phase, while in the temporal gauge there is no position dependence but only a temporal dependence of the coin operator.} \cite{AD16a, AD16b}.

\section{Implementation of the circuits with the classical simulator of IBM's quantum processors}
\label{sec:Implementation}

\subsection{Implementation of the naive and linear-depth circuits}

\begin{figure*}
\hspace{-0.4cm}
\includegraphics[width=6.3cm]{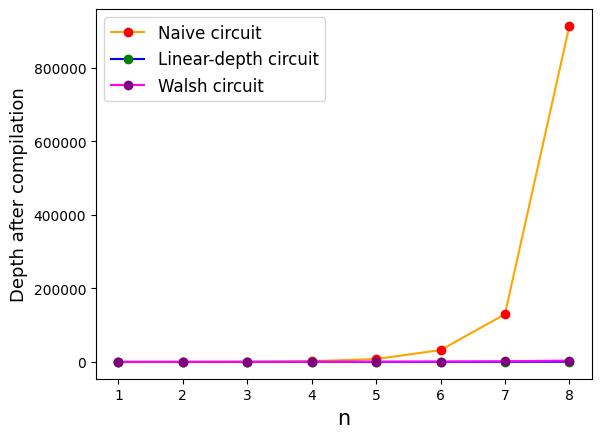} \ \ \ \ \
\includegraphics[width=6cm]{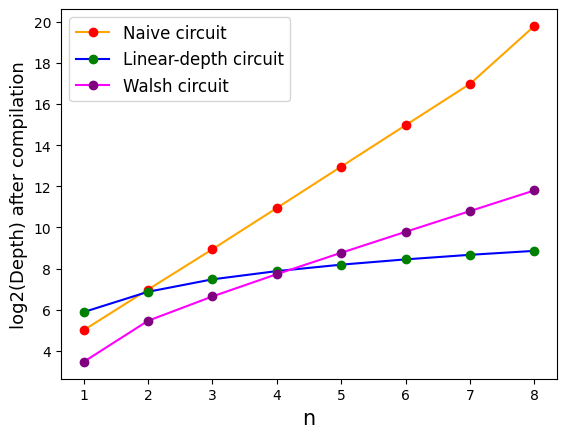} \\
\includegraphics[width=6cm]{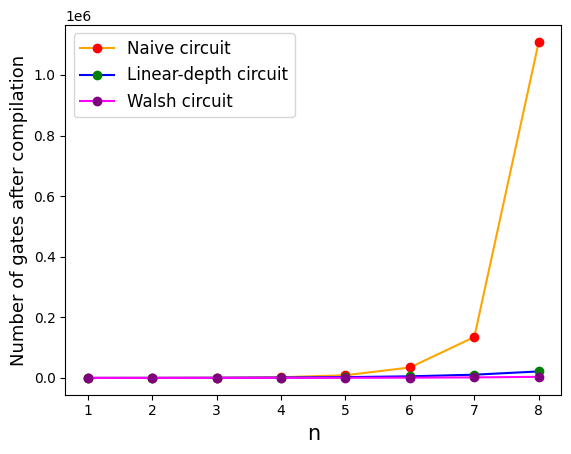} \ \ \ \ \
\includegraphics[width=6cm]{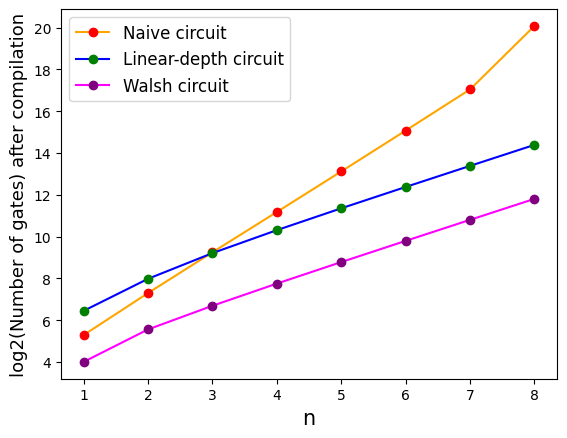}
\caption{Depth (top row) and number of one- and two-qubit gates (bottom row) after compilation of the naive, the linear-depth, and the exact Walsh circuits, as a function of the number $n$ of position qubits (left column), and in log scale (right column). The coin operator chosen for each of the three circuits is Eq.\ \eqref{eq:coin_op_new}, with the values of the angles chosen pseudo-randomly and given by Table \ref{fig:table}. \label{fig:depth_and_gates}}
\end{figure*}

\begin{figure}
\includegraphics[width=8cm]{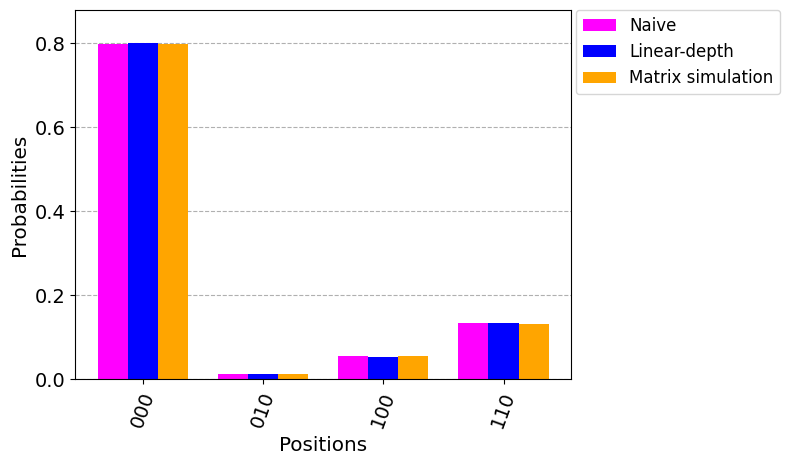}
\caption{Probability distributions obtained by running the naive and the linear-depth circuits on the classical simulator of IBM's quantum processors, over $2^3=8$ nodes, starting from the state $\ket{k_2=000}\ket{s_0=0}$, up to $200$ time steps, and $100000$ times for each circuit. The results are compared to the exact probability distribution, obtained by doing simulations of the DQW directly on the line (without passing by any quantum circuit and base-2 encoding of the position), by matrix multiplications, on Python. The values chosen for the angles defining the coin operators at each node, see Eq.\ \eqref{eq:coin_op_new}, are given in Table \ref{fig:table}. \label{fig:histograms}}
\end{figure}

\begin{figure}
\begin{tabular}{| c | c | c | c| c|}
\hline
 &        $\alpha$ & $\theta$ & $\phi$ & $\lambda$ \\ \hline
 $C_0$ & 2.79345642 & 2.04965065 & -1.40857922 & -0.92367785 \\ \hline
  $C_1$ & 1.06079763 & 0.78803616 & 2.65787445 & -2.41605216 \\ \hline
  $C_2$ & 1.37954004 &  2.33616145 &  1.80543107 & 1.62653295 \\ \hline
  $C_3$ & 2.19632113 & 0.83464054 & -0.6205838 & -2.18954578 \\ \hline
  $C_4$ & 1.68034403 & 0.14646079 & -1.58351372 &  0.15425736 \\ \hline 
  $C_5$ & 1.87405243 &  2.83703533 &  1.81087703 & -2.22138109 \\ \hline 
  $C_6$ & 1.10728697 & 2.5120075 &  0.46211176 &  1.15354949 \\ \hline 
  $C_7$ & 1.97339916 & 0.5405225 &  2.67464114 & -1.52776719 \\ \hline
\end{tabular}
\caption{Values drawn at random for the angles of the coin operators at each position, which we have used to generate the histograms of Fig.\ \ref{fig:histograms}. \label{fig:table}}
\end{figure}

\begin{figure*}
\includegraphics[width=7cm]{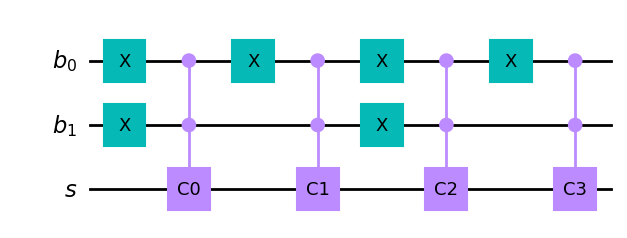}
\includegraphics[width=10cm]{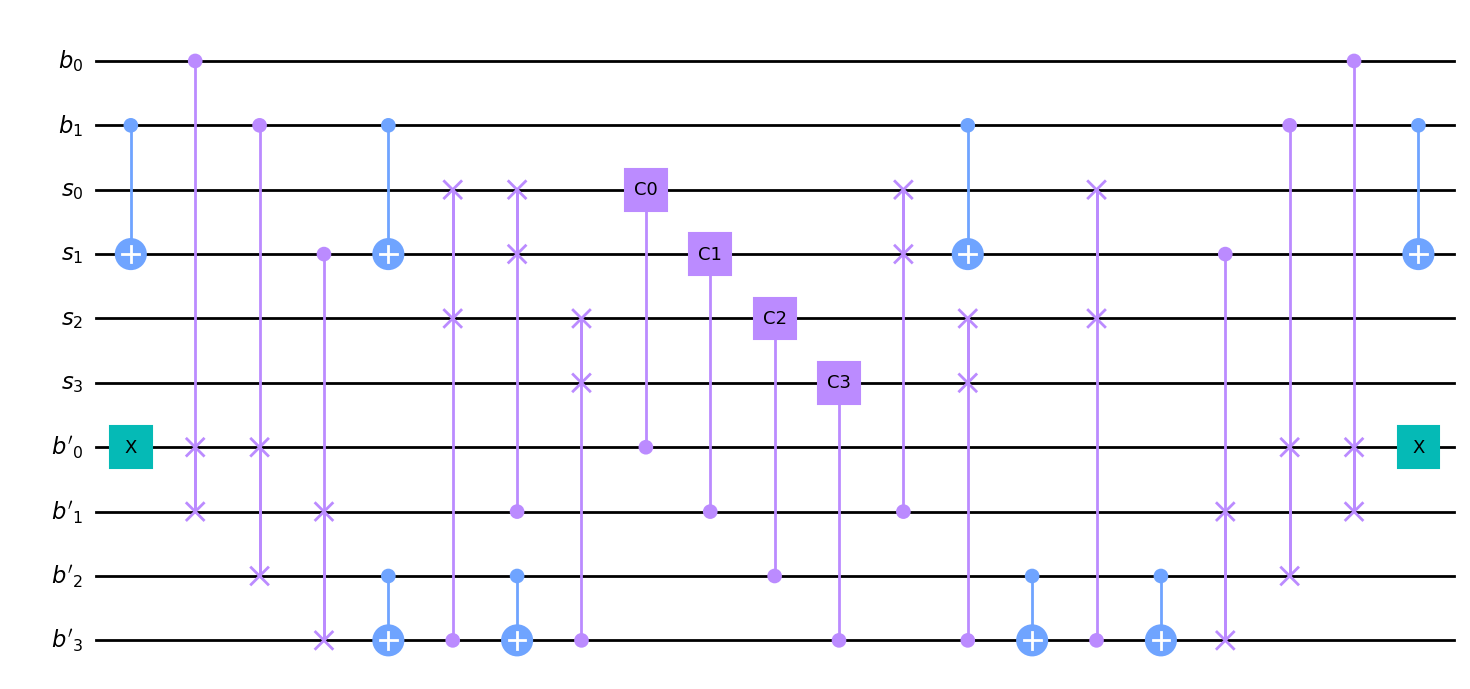}
\caption{Naive (left figure) and linear-depth (right figure) circuits for $n=2$, as generated by Qiskit, before compilation, i.e., as we have defined them. \label{fig:circuits_Qiskit}}
\end{figure*}

We have implemented the naive and the linear-depth circuits on the classical simulator of IBM's quantum processors, called QASM (Quantum Assembly Language) simulator. We have used the software Qiskit to code the circuits.

First of all, we have studied how the depth and the number of gates of the naive and linear-depth circuits vary with the number $n$ of position qubits.
This is shown in Fig.\ \ref{fig:depth_and_gates}.
After compilation, there are only one- and two-qubit gates.
The set of one- and two-qubit types of gates with which we have decomposed the circuits is:
\begin{subequations}
\begin{align}
R_X(\theta) &\defeq
\begin{bmatrix}
\cos \frac{\theta}{2} & -i \sin \frac{\theta}{2} \\
 -i \sin \frac{\theta}{2} & \cos \frac{\theta}{2}
\end{bmatrix} \label{eq:RX}\\
R_Y(\theta) &\defeq
\begin{bmatrix}
\cos \frac{\theta}{2} & - \sin \frac{\theta}{2} \\
 \sin \frac{\theta}{2} & \cos \frac{\theta}{2}
\end{bmatrix} \\
R_Z(\lambda) &\defeq
\begin{bmatrix}
e^{-i \frac{\lambda}{2}} & 0 \\
0 & e^{i \frac{\lambda}{2} }
\end{bmatrix} \\
P(\lambda)  &\defeq
\begin{bmatrix}
1 & 0 \\
0 & e^{i {\lambda}} 
\end{bmatrix} \\
CNOT& \defeq
\begin{bmatrix}
1 & 0 & 0 & 0 \\
0 & 1 & 0 & 0 \\
0 & 0 & 0 & 1 \\
0 & 0 & 1 & 0 
\end{bmatrix} \, .
\end{align}
\end{subequations}

Then, we have made simulations of the DQW (both the naive and the linear-depth circuits) for $n=3$ position qubits, i.e., $2^3=8$ positions, starting with the state $\ket{k_2=000}\ket{s_0=0}$, and up to $200$ time steps (remember that we have periodic boundary conditions), after which we measure the state.
This has been repeated $100 000$ times for each type of circuit (the naive and the linear-depth ones), which delivers the histograms of Fig.\ \ref{fig:histograms}.
The position-dependent coin operator chosen for these simulations is one for which each coin operator at a given position has been generated (pseudo-)randomly.
The parametrization of the coin operator at each position is
\begin{equation}
K(\alpha,\theta,\phi,\lambda) \defeq e^{i\alpha}
\begin{bmatrix}
\cos \frac{\theta}{2} & - e^{i\lambda} \sin \frac{\theta}{2} \\
e^{i\phi} \sin \frac{\theta}{2} &  e^{i(\phi+\lambda)} \cos \frac{\theta}{2} 
\end{bmatrix} \, .
\label{eq:coin_op_new}
\end{equation}
The intervals over which we draw the values of the angles at random are
\begin{equation}
\alpha,\theta \in [0, \pi[ \ \text{and} \ \phi, \lambda \in [-\pi, \pi[ \, ,
\end{equation}
and the values that we have obtained for the angles are given in Table \ref{fig:table} .

Finally, we show in Fig.\ \ref{fig:circuits_Qiskit} the naive and linear-depth circuits coded on Qiskit, before compilation, for $n=2$ position qubits (for $n=3$ the circuits are too large).

\subsection{Implementation of the Walsh circuit}

\begin{figure*}
 \includegraphics[width=5.2cm]{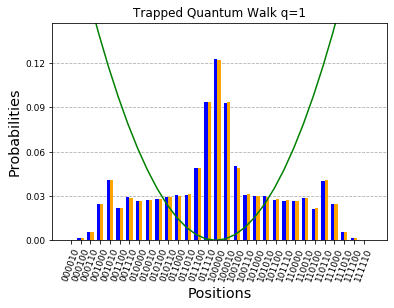}
 \includegraphics[width=5.2cm]{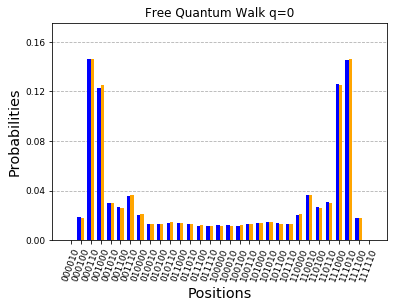}
 \includegraphics[width=7.2cm]{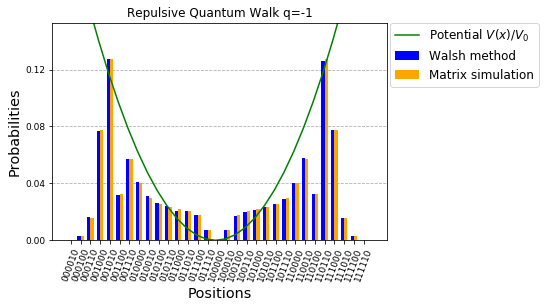}
\caption{Simulations of the Dirac DQW, with walk operator given by Eq.\ \eqref{eq:the_walk_op} and coin operator given by Eq.\ \eqref{eq:Dirac_coin_op}, in a harmonic potential $V(x)=V_0(x-0.5)^2$, after $30$ time steps, implemented on IBM's QASM simulator, using $n=6$ position qubits and an exact Walsh decomposition of the potential $e^{-iqV(\hat{x})a \otimes \hat{1}}$. We have compared this simulation to a standard simulation of the DQW on the line via products of matrices, which we call ``Matrix simulation''. The initial state of the DQW is a peak of probability on the middle position $N/2$, and balanced between the two coin states, that is, $(1/\sqrt{2}) \ket{(N/2)_2} (\ket{\uparrow} + \ket{\downarrow})$. The left figure corresponds to a DQW with a positive charge $q=1$, so that the particle is trapped by the potential. The middle figure corresponds to a free DQW, without potential. The right figure corresponds to a DQW with a negative charge $q=-1$, so that the particle starts at an unstable position, the top of the potential. The other parameters are $n=6, m=10, a =0.05, V_0=80\pi$.
The histograms have been produced by running the DQW and measuring its position 100000 times for each figure. \label{fig:Dirac}}
 \end{figure*}

We have implemented the Walsh-decomposition-based circuit (which we simply call Walsh circuit) on the classical simulator of IBM's quantum processors, called QASM simulator.
We have used the software Qiskit to code the circuit.
The DQW is still Eq.\ \eqref{eq:the_walk_scheme} with the walk operator given by Eq.\ \eqref{eq:the_walk_op}, and we have chosen the following coin operator,
\begin{equation}
\label{eq:Dirac_coin_op}
C^{(n)} \defeq e^{-iqV(\hat{x})a\otimes \hat{1}}R_X(-2ma) \, ,
\end{equation}
where $\hat{x} \equiv \hat{k}a$ is the position operator on the 1D lattice, $a$ is the position and time step, $\hat 1$ is the identity $2\times 2$ matrix, $q$ and $m$ are some real parameters, $R_X$ is given by Eq.\ \eqref{eq:RX}, and $V(x)$ is a real-valued function defined on $[0,1]$ that we have chosen to be harmonic,
\begin{equation}
V(x) \defeq V_0(x-0.5)^2 \, ,
\end{equation}
where $V_0$ is some constant.
One can easily show that this DQW has a continuum-limit description for $a \rightarrow 0$, namely, the following Dirac equation in a potential $V$ (in units $c=1$, $ \hbar =1$),
\begin{equation}
 (i\gamma^{0}(\partial_{t}+iqV)+i\gamma^{1}\partial_{x})\psi=m\psi \, ,
\end{equation}
with  $\gamma^0\defeq \sigma^1$ and $\gamma^1\defeq i\sigma^2$.
Hence, we call this DQW a Dirac DQW, even if we are not in the continuum limit.
Results of these simulations are presented in Fig.\ \ref{fig:Dirac}.
In Fig.\ \ref{fig:depth_and_gates}, we have shown how, for this Walsh circuit, the depth and number of one- and two-qubit gates after compilation varies with the number $n$ of qubits, having chosen for the coin operators the same as those used for the naive and linear-depth circuits in that figure, that is, Eq.\ \eqref{eq:coin_op_new} with the pseudo-random angles given by Table \ref{fig:table}.

\section{Conclusions and discussion}
\label{sec:conclusions}

In this paper we have reported on the quantum-circuit implementation of position-dependent coin operators in DQWs.
In Sec.\ \ref{sec:1}, we have recalled the definition of the DQW, made of the concatenation of a coin-dependent shift operator and a coin operator.
In Sec.\ \ref{sec:shift}, we have recalled the QFT scheme quantum-circuit implementation of the coin-dependent shift operator, presented in Ref.\ \cite{Shakeel2020a}, and demanding less resources than the ID scheme of Ref.\ \cite{FOBH2005}.
Then we have proceeded to the presentation of the results of this work.
In Sec.\ \ref{sec:naive}, we have presented a certain quantum circuit implementing a position-dependent coin operator, the depth of which is exponential in the number of position wires $n$, but which uses no ancillary qubits.
 This is because in this circuit we apply sequentially the coin operators at each position, which is why we call this the naive circuit.
In Sec.\ \ref{sec:linear}, we present a new quantum circuit implementing a position-dependent coin operator, the depth of which is this time linear in $n$, at the cost of introducing an exponential number of ancillary wires.
The main idea of this linear-depth circuit is to implement in parallel (rather than sequentially) all coin operators at the different positions.
The linear-depth circuit is made of three blocks: we first initialize ancillary wires with the information about the position of the walker, we then initialize other ancillary wires with the information about the coin state at each position, and we finally apply all coin operators in parallel, after which we undo both initializations.
Finally, in Sec.\ \ref{sec:efficient}, we have extended the result of Ref.\ \cite{WGMAG2014} from position-dependent unitaries which are diagonal in the position basis to position-dependent \emph{block-diagonal} unitaries.
Indeed,  we have shown that for a position dependence of the coin operator (the block-diagonal unitary) which is smooth enough, one can find an efficient quantum-circuit implementation approximating the coin operator up to an error $\epsilon$ (in terms of the spectral norm), the depth and size of which scale as $O(1/\epsilon)$.
In Sec.\ \ref{sec:Implementation}, we have implemented the naive, the linear-depth, and the Walsh circuits, on the classical simulator of IBM's quantum processors, to give an idea of the results one obtains for a small number of position qubits.

We are currently working on a circuit the depth of which is tunable in an exponential manner, in between an exponential depth and a linear depth: that is, if the depth of the naive circuit is proportional to $2^n$ where $n$ is the number of position wires, then we want that a depth with any power $2^p$ of $2$ be achievable by our new circuit, at the cost of introducing the right amount of ancillae, which would grow exponentially as the depth diminishes exponentially.
This is reminiscent of Ref.\ \cite{APLOplus2021} for quantum state preparation, where a trade-off between the depth and the ancillae is proposed.

Let us now mention some possible applications of our findings.
During the NISQ era, an efficient design of the quantum circuit implementing the desired DQWs is crucial in order to make use of their potential as quantum simulators.
Our proposed \emph{linear-depth implementation} of the position-dependent coin operator, as well as the \emph{efficient implementation} of the smooth position-dependent coin operator, constitute a step forward in this direction, and pave the way to simulating many interesting physical scenarios.
A first example of application of our circuits is the implementation of DQWs coupled to various gauge fields, either electromagnetic \cite{DDMEF12a, AD16a, AD16b, MMAMP18, CGWW18}, non-Abelian Yang-Mills \cite{AMBD16}, or gravitational \cite{DMD13b, DMD14, AD17, AF17}, for which the efficient implementation is likely to be possible.
Notice that recently, quantum spatial-search schemes have been developed which use gauge fields as the oracle, to mark the vertex to be found \cite{ZD2021, FZAD2022}.
A second example is the implementation of DQWs with certain random coin operators, which leads to diffusion \cite{DiMolfetta2016, AMACplus20}.
A third example is that of DQWs leading to localization, either when spatial disorder is included \cite{JM2010, SCPGplus2011, CORGplus2013}, or due to non-linear effects \cite{NBPR2010} (where soliton-like structures appear), or by the use of a spatially periodic coin operator \cite{SK2010}.
Notice that in the case of spatial disorder, i.e., spatial noise, on the coin operator, the efficient implementation will not be possible anymore, but we can use the linear-depth one.
Interestingly enough, localization can also appear as a consequence of the interaction with a smooth external potential, instead of a random, or even periodic perturbation: indeed, such a localization has been evidenced in the DQW simulation of a spin-1/2 particle in extra physical dimensions \cite{MMDMplus2017, ACP2022}.
Finally, neutrino oscillations in matter can also be simulated with the help of DQWs with a position-dependent coin operator \cite{DMP2016}.
Most probably, the above examples do not cover all the possibilities that will appear in the future.

Finally, let us comment on the experimental requirements of the circuits we propose in this article.
Let us start with general comments.
In order to realize many steps of the DQW, one needs a large scalability of the experimental platform, i.e., one needs the capacity to conceive an experimental setup that allows for exploring a large number of steps, i.e., a graph with many nodes (in this work, the graph is the 1D line, with a given number of nodes on this line).
Moreover, for the experimental final probability distribution to match the theoretical one, one needs noise to be as reduced as possible.
Let us now give more precision on the noise requirements.
Let us consider that we want to implement experimentally the DQW of this paper over 8 nodes on the line.
What is the number of one- and two-qubit gates that one needs in order to implement an arbitrary position-dependent coin operator over 8 nodes (for an arbitrary initial state)?
We recall that this is the number of gates after the compilation, which has been done with the gates $R_X$, $R_Y$, $R_Z$, $P$ and $CNOT$.
For the linear-depth circuit, this number of gates is $N_{\text{L}}^{\text{coin}}(8) = 591$;
this gate count has been obtained for a position-dependent coin operator as arbitrary as possible, that is, of the form of Eq.\ \eqref{eq:coin_op_new} with pseudo-random values for each of the four angles, given in Table \ref{fig:table}.
For the Walsh circuit, this number of gates is, for a null error between the circuit we implement and the exact position-dependent coin operator, $N_{\text{W}}^{\text{coin}}(8) = 103 $; this gate count has been obtained for a position-dependent coin operator as arbitrary as possible, that is, of the form of Eq.\ \eqref{eq:arbitrary_coin_op} with pseudo-random values for each of the four angles\footnote{It has been checked that $N_{\text{W}}^{\text{coin}}(8) $ stays roughly the same whatever pseudo-random values we choose for the angles. Also, let us mention that we could as well have chosen the parametrization of Eq.\ \eqref{eq:coin_op_new}.}.
For $k$ time steps, the number of gates involved by the position-dependent coin operators solely is thus $k \times N_i^{\text{coin}}(8)$, for $i = \text{L}, \text{W}$.
Moreover, let $N^{\text{shift}}(8)$ be the number of gates needed to implement the shift operator over 8 nodes with the quantum-Fourier-transform scheme.
We have that $N^{\text{shift}}(8) = 30 $.
Let $\varepsilon$ be the total, final error between the theoretical and the experimental implementations of the circuits.
Since the total error of a circuit is the addition of the errors of each one- and two-qubit gate \cite{book_nielsen_chuang_2010}, the error per gate $\varepsilon_{\text{gate},i}$ that we can allow for a total error of $\varepsilon$ is $\varepsilon_{\text{gate},i} = \varepsilon / [k (N_i^{\text{coin}}(8) + N^{\text{shift}}(8)) ]$.
For $k=4$ and $\varepsilon = 1\% = 0.01$, this gives $\varepsilon_{\text{gate},\text{L}} = 4.025 \times 10^{-6} $ and $\varepsilon_{\text{gate},\text{W}} = 18.80 \times 10^{-6} $.
These error requirements are several orders of magnitude below current standards in quantum computers, although the quality of qubits is continuously improving.

\begin{center}

{\bfseries AUTHOR CONTRIBUTIONS}

\end{center}

U. Nzongani and C.-E. Doncecchi have found the naive and the linear-depth quantum circuits.
J. Zylberman has extended the result of Ref.\ \cite{WGMAG2014}, providing the approximate efficient quantum-circuit implementation for a $2\times 2$-block-diagonal unitary (the coin operator).
P. Arnault has provided all the proofs of the appendices for the naive and the linear-depth circuits (not Appendix \ref{app:EQC}), and has handled the writing of the whole manuscript.
A. P{\'e}rez has written Sec.\ \ref{sec:shift} on the quantum circuits available to implement the shift operator of the DQW.
F. Debbasch has participated to several discussions, and provided his insight on the final manuscript.

\begin{center}

{\centering \bfseries{DATA AVAILABILITY}}

\end{center}

Data will be made available upon reasonable request.

\begin{center}

{\bfseries STATEMENT OF ABSENCE OF CONFLICT OF INTEREST}

\end{center}

On behalf of all authors, the corresponding authors state that there is no conflict of interest.

\begin{center}

{\centering \bfseries{ACKNOWLEDGEMENTS}}

\end{center}

The authors thank T. Fredon for stimulating discussions.
This work has been supported by the PEPR integrated project EPiQ ANR-22-PETQ-0007, part of Plan France 2030.
On A. Perez’ side, this work has been funded by: 
the Spanish MCIN/AEI/10.13039/501100011033 grant PID2020-113334GB-I00; 
the MINECO Grant SEV-2014-0398;
the Generalitat Valenciana grant PROMETEO/2019/087;
the Ministry of Economic Affairs and Digital Transformation of the Spanish Government through the “QUANTUM ENIA project call – Quantum Spain project”;
the European Union, through the “Recovery, Transformation and Resilience Plan – NextGenerationEU” within the framework of the “Digital Spain 2025 Agenda”;
the CSIC Interdisciplinary Thematic Platform (PTI+) on Quantum Technologies (PTI-QTEP+).


%

\appendix

\section{Fujiwara et al.'s scheme, also called increment and decrement (ID) scheme, for the shift operator}
\label{app:Fuji}

\begin{figure}[t]
	\includegraphics[width=8.5cm]{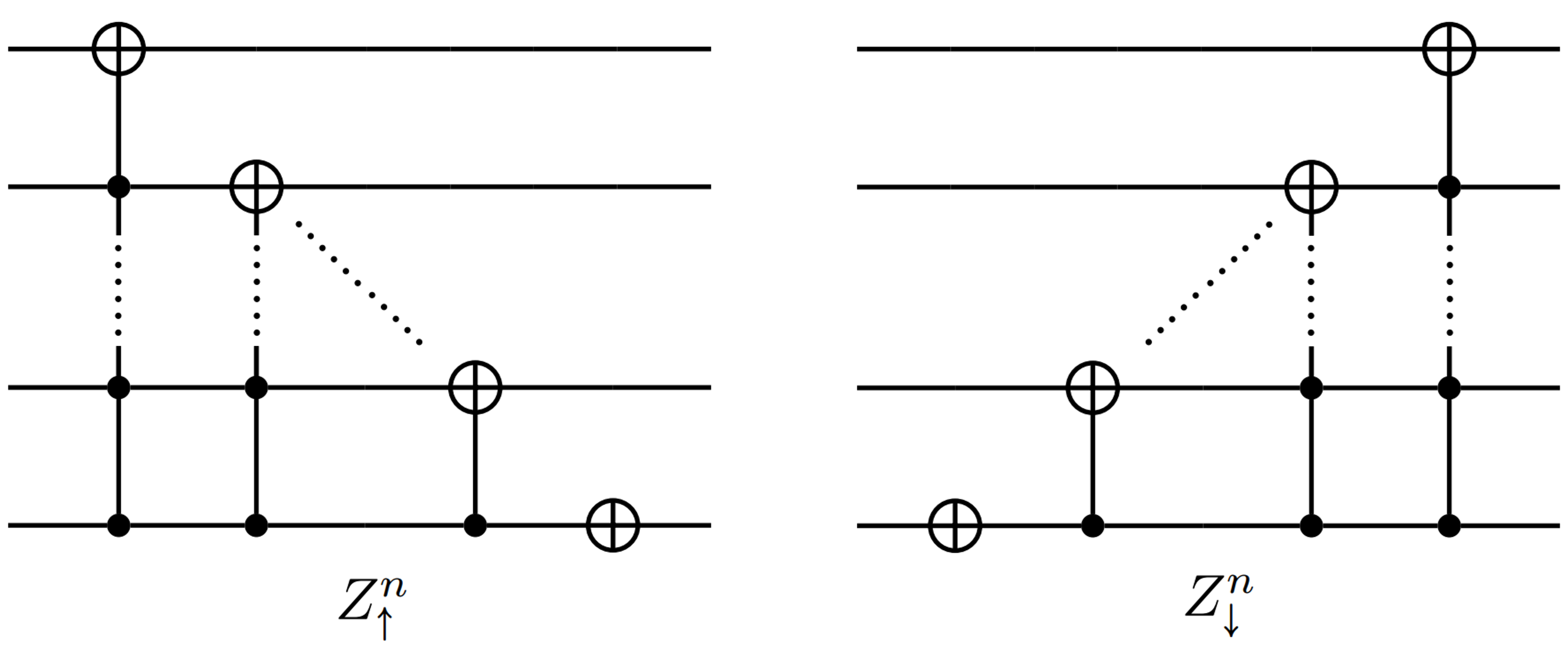}
	\caption{Circuits implementing the left (left figure) and right (right figure) shift operators, respectively $Z^n_{\uparrow}$  and $Z^n_{\downarrow}$, defined respectively in Eqs.\ \eqref{eq:Znup} and \eqref{eq:Zndown}, introduced in Ref.\ \cite{FOBH2005}. \label{fig:Zs}}
\end{figure}

In the case of a uniform total coin operator, i.e., when for each $k$, $C_k = C$, then the total coin operator given by Eq.\ \eqref{eq:non-uniform_coin_operator} reads
\begin{equation}
	{C}^{(n)}_{\text{unif.}} = I_{2^n} \otimes C \, .
\end{equation}

What is proven in Ref.\ \cite{FOBH2005} is that ${W}^{(n)}$ can be implemented by the following circuit,
\begin{equation}
	\label{eq:uniform_walk}
	{W}^{(n)} =        D^{(n)} {C}^{(n)}_{\text{unif.}} \, ,
\end{equation}
since the displacement operator $D^{(n)}$, also called increment and decrement (ID) scheme, can be shown to be equal to the coin-dependent shift operator ${S}^{(n)}$ (see Ref.\ \cite{FOBH2005}), i.e.,
\begin{equation}
	D^{(n)} =  {S}^{(n)} \, ,
\end{equation}
and it is defined by
\begin{equation}
	D^{(n)} \defeq K_1^{n}(Z^{n}_{\uparrow}) (I_{2^n} \otimes X) K_1^{n}(Z^{n}_{\downarrow}) (I_{2^n} \otimes X) \, ,
\end{equation}
where $K_m^{m'}(M)$ is the application of the gate $M$ occupying  $m'$ wires and being controlled by the first $m$ wires starting from wire $0$, where $X = \sigma^1 = \begin{bmatrix}
0 & 1 \\ 1 & 0
\end{bmatrix}$, and where the left and right displacement operators are respectively defined by 
\begin{subequations}
	\begin{align}
		Z^{n}_{\uparrow} &\defeq \sideset{}{^L}\prod_{p=0}^{n-1} I_{2^{n-1-p}} \otimes K_p^{1}(X) 
		\label{eq:Znup} \\
		Z^{n}_{\downarrow} &\defeq  \sideset{}{^R}\prod_{p=0}^{n-1} I_{2^{n-1-p}} \otimes K_p^{1}(X)  \, ,
		\label{eq:Zndown}
	\end{align}
\end{subequations}
where the superscripts $L$ and $R$ mean respectively that the product is performed from right to left (and hence towards the left) or from left to right (and hence towards the right).

\begin{figure}[t]
	\includegraphics[width=8.5cm]{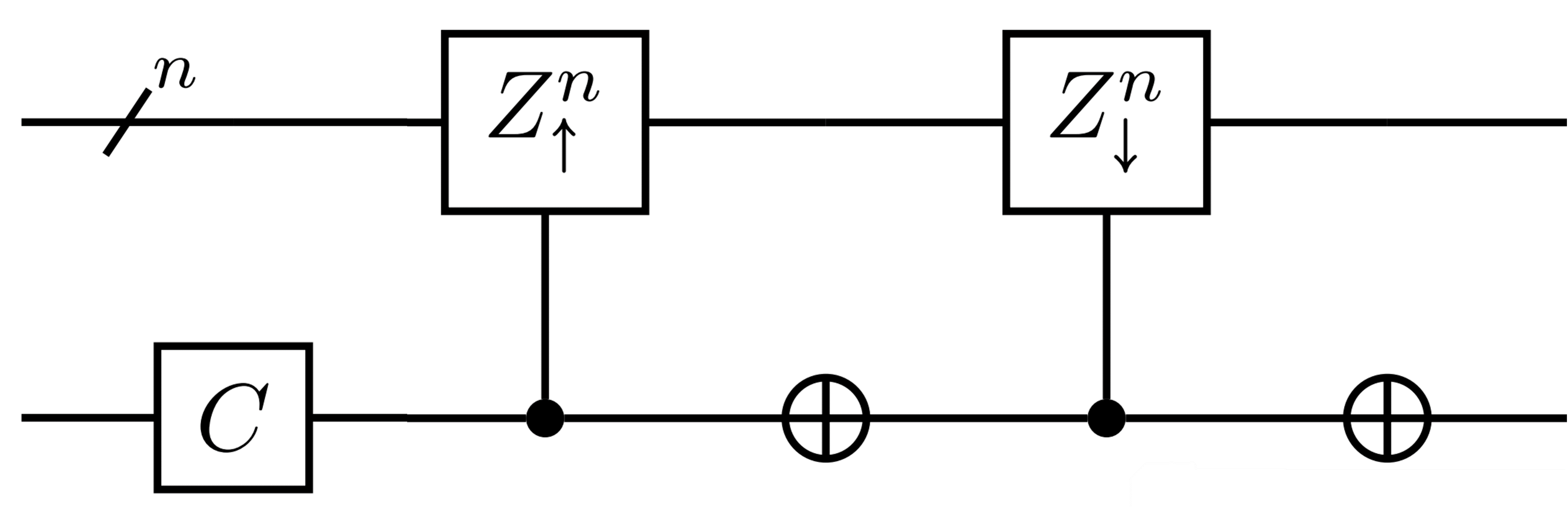}
	\caption{Circuit implementing the uniform discrete-time quantum walk characterized by the walk operator of Eq.\ \eqref{eq:uniform_walk}, introduced in Ref.\ \cite{FOBH2005}.	\label{fig:DC} }
\end{figure}

In Figure \ref{fig:Zs} we show the circuits implementing the left and right shift operators, respectively $Z^n_{\uparrow}$ and $Z^n_{\downarrow}$, defined respectively Eqs.\ \eqref{eq:Znup} and \eqref{eq:Zndown}, and in Fig.\ \ref{fig:DC} we show the circuit implementing the uniform DQW characterized by the walk operator of Eq.\ \eqref{eq:uniform_walk}.

\section{Proof of $U^{(n)} = {C}^{(n)}$, by induction on $n$}
\label{app:proof}

It is easy to show that $U^{(1)} = {C}^{(1)}$.
Let us now show that assuming $U^{(n)} = {C}^{(n)}$ for a certain $n$ implies $U^{(n+1)} = {C}^{(n+1)}$.
We start by expressing $U^{(n+1)}$, given by Eq.\ \eqref{eq:the_circuit}, as the product of two circuits,
\begin{equation}
\label{eq:decomposition}
U^{(n+1)} = P^{(1)}_{n+1} P^{(0)}_{n+1} \, ,
\end{equation}
where
\begin{equation}
\label{eq:subcircuit}
P^{(a)}_{n+1} \defeq  \sideset{}{^L}\prod_{k=0}^{2^n-1} G_{n+1}(C_{2^na + k}) T_{n+1}(k) \, . 
\end{equation}
Considering the definition of the towers $T_n$ given in Eq.\ \eqref{eq:towers}, it makes sense to write $P^{(a)}_{n+1}$ given in Eq.\ \eqref{eq:subcircuit} as
{\small
\begin{equation}
\label{eq:equation}
P_{n+1}^{(a)} = \Big\{ \, \sideset{}{^L}\prod_{k=1}^{2^n-1} G_{n+1}(C_{2^na + k}) T_{n+1}(k) \Big\} \Big\{ G_{n+1}(C_{2^na}) T_{n+1}(0)  \Big\}
\end{equation}}
Now, using the definitions of Eqs.\ \eqref{eq:towers} and \eqref{eq:Gs}, we can rewrite Eq.\ \eqref{eq:equation} as
\begin{widetext}
\begin{subequations}
\begin{align}
P_{n+1}^{(a)} &= \Bigg\{ \, \sideset{}{^L}\prod_{k=1}^{2^n-1} \begin{bmatrix}
I_{2^{n+1}} & 0 \\ 0 & G_n(C_{2^n a + k})
\end{bmatrix} 
\begin{bmatrix}
T_n(k \ \text{mod} \ 2^{n-1} ) & 0 \\ 0 & T_n(k \ \text{mod} \ 2^{n-1} )
\end{bmatrix}
\Bigg\}
\Bigg\{
\begin{bmatrix}
I_{2^{n+1}} & 0 \\ 0 & G_n(C_{2^n a})
\end{bmatrix} 
\begin{bmatrix}
0 & T_n(0) \\ T_n(0) & 0
\end{bmatrix} 
\Bigg\} \\
&= 
\Bigg\{ \, \sideset{}{^L}\prod_{k=1}^{2^n-1} \begin{bmatrix}
T_n(k \ \text{mod} \ 2^{n-1} ) & 0 \\ 0 & G_n(C_{2^n a + k}) T_n(k \ \text{mod} \ 2^{n-1} )
\end{bmatrix} 
\Bigg\}
\begin{bmatrix}
0 & T_n(0) \\ G_n(C_{2^n a}) T_n(0) & 0
\end{bmatrix} \, . \label{eq:second}
\end{align}
\end{subequations}
\end{widetext}
Now, it is easy to show the following formula,
{\small
\begin{equation}
\begin{bmatrix}
A_N & 0 \\ 0 & B_N
\end{bmatrix}
\cdots  \begin{bmatrix}
A_1 & 0 \\ 0 & B_1
\end{bmatrix}
\begin{bmatrix}
0 & D \\ C & 0
\end{bmatrix} =
\begin{bmatrix}
0 & A_N ... A_1 D \\ B_N ... B_1 C & 0
\end{bmatrix} \, ,
\end{equation}}
which, applied to Eq.\ \eqref{eq:second}, leads to
\begin{equation}
\label{eq:Pn}
P^{(a)}_{n+1} = 
\begin{bmatrix}
0 & Q_n \\ R_n^{(a)} & 0
\end{bmatrix} \, ,
\end{equation}
with
\begin{subequations}
\begin{align}
Q_n &\defeq  \sideset{}{^L}\prod_{k=0}^{2^n-1} T_n( k \ \text{mod} \ 2^{n-1} ) \\
R_n^{(a)} &\defeq   \sideset{}{^L}\prod_{k=0}^{2^n-1} G_n(C_{2^na+k}) T_n( k \ \text{mod} \ 2^{n-1} ) \, .
\end{align}
\end{subequations}
We immediately notice, see Eq.\ \eqref{eq:the_circuit}, that
\begin{equation}
R_n^{(0)} = U^{(n)} \, ,
\end{equation}
which by induction assumption is equal to ${C}^{(n)}$, and we also have that
\begin{equation}
R_n^{(1)} = \sideset{}{^L}\prod_{k=0}^{2^n-1} G_n(C_{2^n+k}) T_n( k \ \text{mod} \ 2^{n-1} )  \, ,
\end{equation}
so that we can also use the induction assumption and obtain
\begin{equation}
R_n^{(1)} = {C}^{(n)}_{\text{bis}} \, ,
\end{equation}
with
\begin{equation}
{C}^{(n)}_{\text{bis}} \defeq \sum_{k=2^n}^{2^{n+1}-1} \ket{k_2} \! \! \bra{k_2} \Big|_{\mathcal{S}} \otimes \tilde{C}_{k_2} =
\begin{bmatrix}
C_{2^n} & & 0 \\
 & \ddots &  \\
 0 & & C_{2^{n+1}-1}
\end{bmatrix} \, ,
\end{equation}
where $\ket{k_2} \! \! \bra{k_2} \Big|_{\mathcal{S}}$ means that we consider the operator $\ket{k_2} \! \! \bra{k_2}$ restricted to the subspace $\mathcal{S}$ formed by the vectors $\ket{k_2}$ for $k_2=2^n, ...,2^{n+1}-1$.
Moreover, it is easy to show by induction on $n$ that
\begin{equation}
Q_n = I_{2^n} \, .
\end{equation}
Hence, using Eq.\ \eqref{eq:Pn} for $a=0$ and $1$ in Eq.\  \eqref{eq:decomposition} leads to
\begin{subequations}
\begin{align}
U^{(n+1)} &= \begin{bmatrix}
0 &  I_{2^n} \\ {C}^{(n)}_{\text{bis}}  & 0
\end{bmatrix}
\begin{bmatrix}
0 &  I_{2^n} \\ {C}^{(n)}  & 0
\end{bmatrix} \\
&= \begin{bmatrix}
{C}^{(n)} & 0 \\ 0  & {C}^{(n)}_{\text{bis}}
\end{bmatrix} \\
&= {C}^{(n+1)} \, ,
\end{align}
\end{subequations}
which completes the proof.

\section{Initializing the ancillary positions: $Q_1$}
\label{app:Q1}

\subsection{Basic requirement on $Q_1$}

Let $\ket{K}$ be a state of the following form,
\begin{equation}
\label{eq:K}
\ket{K} \defeq \ket{k_2} \ket{s_0} \ket{s'=0} \ket{b'=0} \, .
\end{equation}
We want a unitary operator $Q_1$ which acts on states $\ket K$ of the form of Eq.\ \eqref{eq:K} as follows,
\begin{equation}
\label{eq:Q1action}
\begin{split}
  &Q_1 \left(\ket{k_2} \ket{s_0} \ket{s'=0} \ket{b'=0} \right) \\
& \ \ \ \ \ \ \ \ \ \ = \ket{k_2} \ket{s_0} \ket{s'=0} \ket{(2^{k})_2}  \, ,
\end{split}
\end{equation}
where we have used the notation
\begin{equation}
\ket{(2^{k})_2} \defeq  | \underbrace{0...0(b'_k=1)0...0 }_{ 2^n \  \text{locations}} \rangle   \, .  
\end{equation}
In simple words, the action of $Q_1$ on such a state $\ket K$ is to change the $k$th bit of $\ket{b'}$ from $b'_k=0$ to $b'_k=1$.

\subsection{Naive scheme}

\begin{figure*}[t!]
\hspace{-0.5cm}
	\includegraphics[width=5cm]{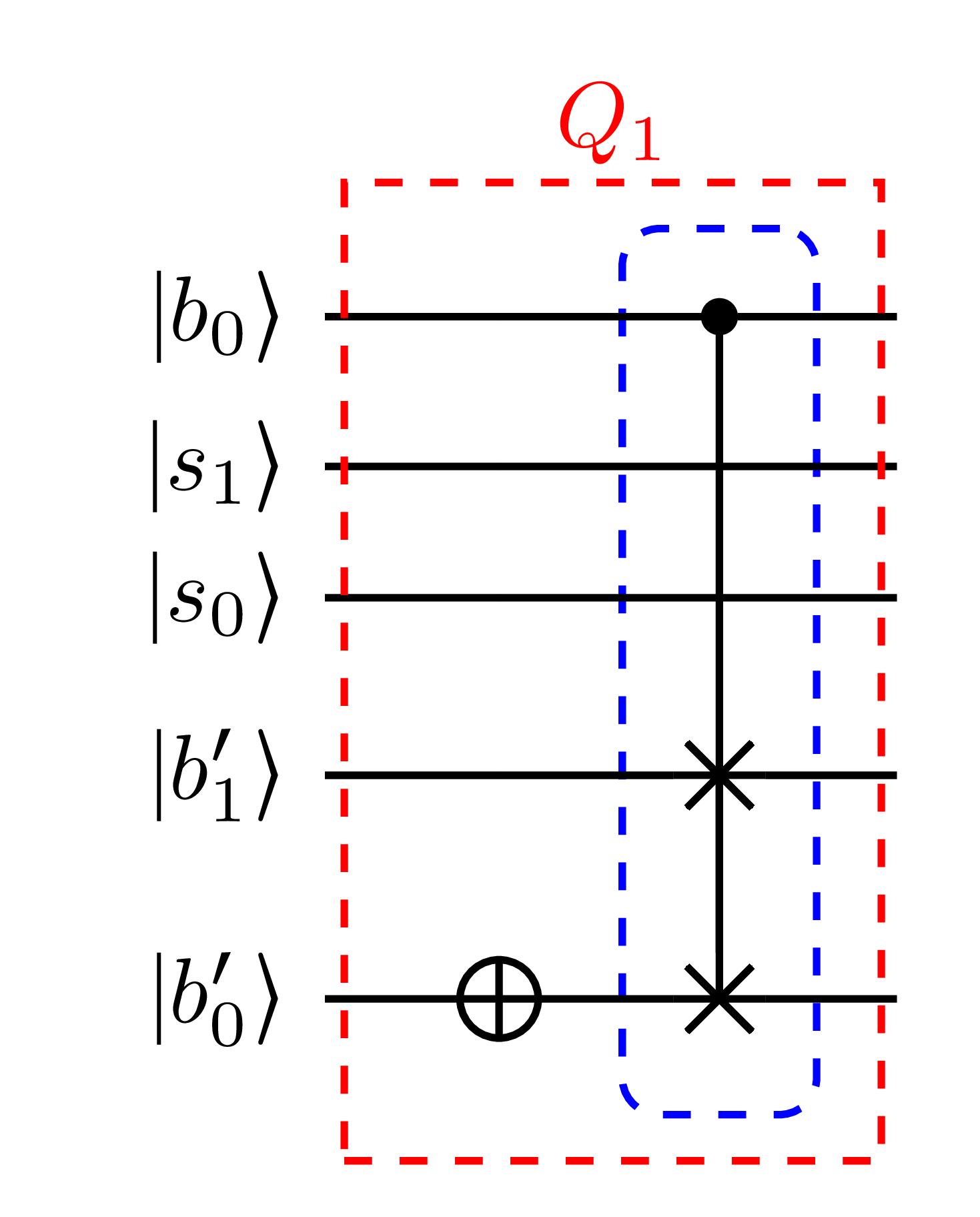}
	\includegraphics[width=5cm]{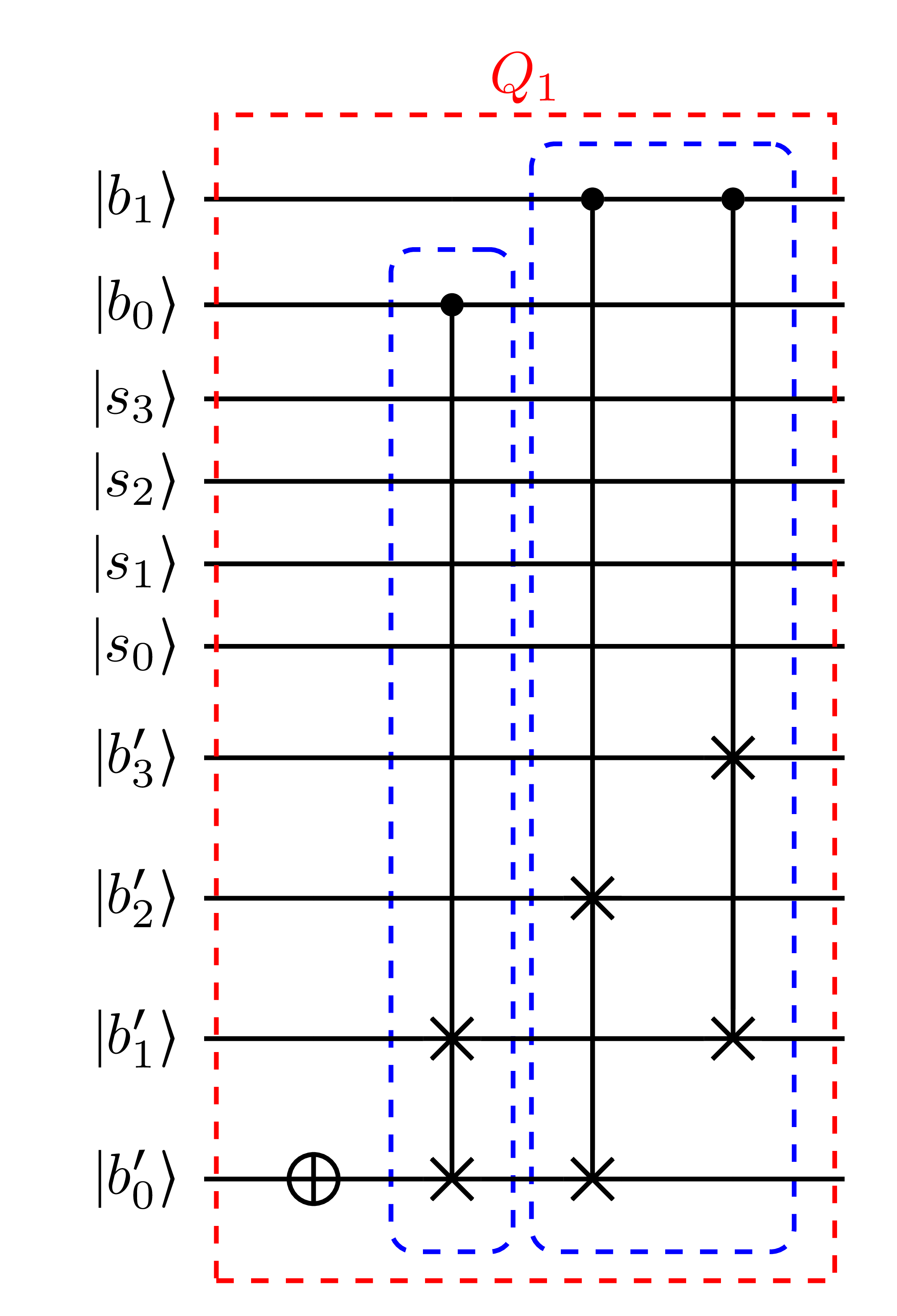}
	\includegraphics[width=7cm]{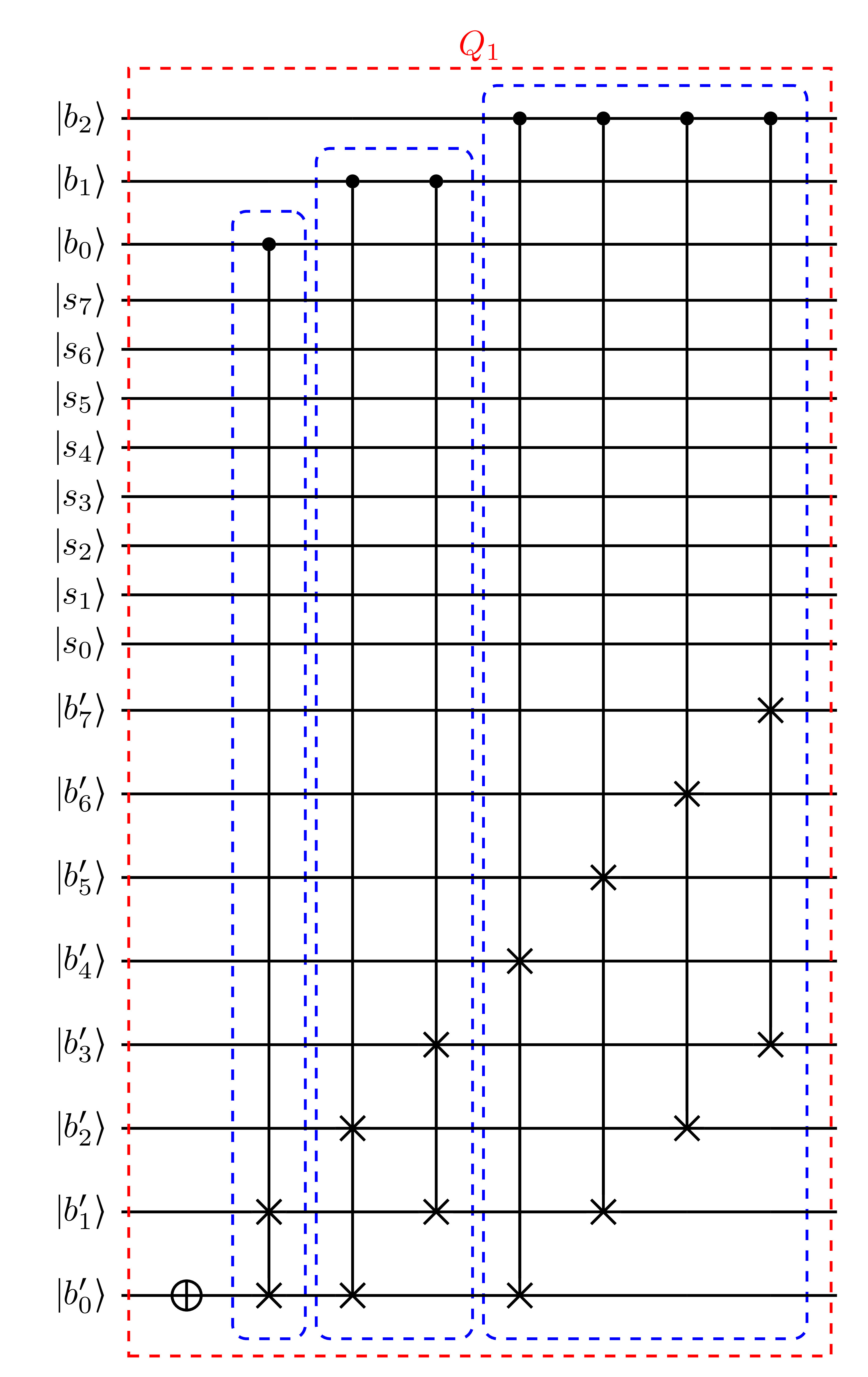}
	\caption{Diagrammatic representation of the naive scheme for $Q_1$ for $n=1$ (left figure), $n=2$ (middle figure), and $n=3$ (right figure). In this naive scheme, all operations inside a same blue box cannot be executed in parallel (except of course for the depth-1 box), which makes the circuit depth exponential in the number of position wires. We remedy this problem by parallelizing these circuits, see Fig.\ \ref{fig:Q1parallelized}. \label{fig:Q1naive}}
\end{figure*}

We want an operator $Q_1$ that is unitary and that satisfies Eq.\ \eqref{eq:Q1action}; we call this a \emph{valid} $Q_1$.
We are going to prove that the circuit depicted in Fig.\ \ref{fig:Q1naive} for $n=1,2,3$, is a valid $Q_1$ for any $n$.

\subsubsection{Definition / Construction}

We use the notation
\begin{equation}
\mathbf{Q}^{(n)}_1 \defeq {Q}_1 \, ,
\end{equation}
which makes explicit the dependence in $n$.
We define $\mathbf{Q}^{(n)}_1$ by induction as follows:
\begin{subequations}
\label{eqs:Q1}
\begin{align}
\mathbf{Q}^{(1)}_1 &\defeq E^{b_0}_{b'_0,b'_1} (I_{2} \otimes I_{2} \otimes I_2 \otimes I_2 \otimes X) 
\label{eq:Q11}\\
\mathbf{Q}^{(n+1)}_1 &\defeq B^{(n+1)} \mathbf{Q}^{(n)}_1 \, ,
\label{eq:induction}
\end{align}
\end{subequations}
with
\begin{equation}
\label{eq:Bnplus1}
B^{(n+1)} \defeq \sideset{}{^L}\prod_{i=0}^{2^n-1}  E^{b_n}_{b'_i,b'_{i+2^n}} \, .
\end{equation}
In Eqs.\ \eqref{eq:Q11} and \eqref{eq:Bnplus1}, we have introduced the controlled exchange (i.e., SWAP) operators $E^{b_n}_{b'_i,b'_{i+2^n}}$, that swap $b'_i$ and $b'_{i+2^n}$ controlling on $b_n$.
In Eqs.\ \eqref{eq:Q11} and \eqref{eq:Bnplus1}, we have purposely omitted to multiply the  $E^{b_n}_{b'_i,b'_{i+2^n}}$ operators by the appropriate identity tensor factors, in order to lighten the notations.
In Eq.\ \eqref{eq:induction}, we have also purposely omitted to multiply $\mathbf{Q}^{(n)}_1$ by the appropriate identity tensor factors, again in order to lighten the notations.

\subsubsection{Proof}

Let us prove by induction on $n$ that the operator $\mathbf{Q}^{(n)}_1$ defined by Eqs.\ \eqref{eqs:Q1} is a valid $Q_1$ operator, i.e., is unitary and satisfies Eq.\ \eqref{eq:Q1action}.
Unitarity is trivial by induction, let us then show the main validity condition.

Let us start by $n=1$, see Eq.\ \eqref{eq:Q11}.
Let us also start with a state $\ket K$ of the form of Eq.\ \eqref{eq:K}, and let us apply $\mathbf{Q}^{(1)}_1$.
After the $X$ gate on $\ket{b_0'}$, we have that $b'_0 = 1$.
Now, if $k=b_0=0$, then the SWAP operation $E^{b_0}_{b'_0,b'_1}$ acts as the identity, and we stay with $b'_0=1$ and $b'_1=0$, which is what we want, i.e., $\ket{b'=(2^k)_2}=\ket{01}$.
But, if $k=b_0=1$, then the SWAP operation $E^{b_0}_{b'_0,b'_1}$ does swap $b'_0$ and $b'_1$, so that we obtain $b'_0=0$ and $b'_1=1$, which is what we want, i.e., $\ket{b'=(2^k)_2}=\ket{10}$.
This completes the proof for $n=1$.

Let us now assume that Eq.\ \eqref{eq:Q1action} is true for $Q_1 = \mathbf{Q}^{(n)}_1$, and let us show that this implies that it is also true for $Q_1 = \mathbf{Q}^{(n+1)}_1$.
There are two situations to be distinguished: either (i) $k \leq 2^n-1$, or (ii) $k\geq 2^n$.
Let us consider situation (i).
We have that
\begin{subequations}
\begin{align}
\mathbf{Q}^{(n+1)}_1 \ket K &\equiv B^{(n+1)} \mathbf{Q}^{(n)}_1  \ket K 
\label{eq:first}\\
&= B^{(n+1)} \left( \ket{k_2} \ket{s_0}\ket{s'=0} \ket{(2^k)_2}\right) \, ,
\label{eq:secondbis}
\end{align}
\end{subequations}
where the first equality \eqref{eq:first} holds by definition of $\mathbf{Q}^{(n+1)}_1$ in Eq.\ \eqref{eq:induction}, and the second equality \eqref{eq:secondbis} holds by induction assumption (it is trivial that this assumption holds when augmenting the Hilbert space from $n$ to $n+1$).
We thus now have to prove that $B^{(n+1)}$ acts as the identity (because the state $\ket{k_2} \ket{s_0}\ket{s'=0} \ket{(2^k)_2}$ is already what we want): this is trivial because all operations in  $B^{(n+1)}$ are controlled by $b_n$, but since $k\leq 2^n-1$ we have that $b_n=0$, so that $B^{(n+1)}$ indeed acts as the identity, which completes the proof for $k\leq 2^n-1$.

Let us now treat situation (ii), $k\geq 2^n$.
We have that
\begin{equation}
k = \underbrace{1}_{=b_n} \times 2^n + l \, ,
\end{equation}
where
\begin{equation}
l \defeq b_{n-1} 2^{n-1} + \dots + b_1 2^1 + b_0 \, .
\end{equation}
Now, $\mathbf{Q}^{(n)}_1$ acting on the appropriate total Hilbert space for $n+1$ is blind to what is on $\ket{b_n}$, and so it simply does by induction assumption what it would do on the Hilbert space for $n$, that is, it encodes $\ket{l}$ on the ancillary positions, that is,
{\small
\begin{subequations}
\begin{align}
\mathbf{Q}^{(n+1)}_1 \ket{K} &\equiv B^{(n+1)} \mathbf{Q}^{(n)}_1  \left( \ket{(2^n+l)_2} \ket{s_0}\ket{s'=0} \ket{(b'=0} \right)
\label{eq:first2}\\
&= B^{(n+1)} \left( \ket{(2^n+l)_2} \ket{s_0}\ket{s'=0} \ket{(2^l)_2}\right) \, ,
\label{eq:second2}
\end{align}
\end{subequations}}
where the first equality \eqref{eq:first2} holds by definition of $\mathbf{Q}^{(n+1)}_1$ in Eq.\ \eqref{eq:induction}, and the second equality \eqref{eq:second2} holds by induction assumption as we have just explained above.
We now have to prove that $B^{(n+1)}$ swaps the $b'_l=1$ that is on the ancillary position $l$ in $\ket{b'}$ with $b'_{2^n+l}=0$.
But this is almost immediate since because $k\geq 2^n$ we have that $b_n=1$ so all the SWAP operations of $B^{(n+1)}$ do swap the relevant qubits, that is, by definition, they scan one by one the $b'_i$s ($i \leq 2^n-1$) and swap them with their corresponding $b'_{i+2^n}$, an operation which is the identity since all ancillary qubits are $0$ except precisely for $b'_l=1$: hence, Eq.\ \eqref{eq:second2} delivers
\begin{equation}
\begin{split}
 &B^{(n+1)} \left( \ket{(2^n+l)_2} \ket{s_0}\ket{s'=0} \ket{(2^l)_2}\right) \\
 & \ \ \ \ \ \ =  \ket{(2^n+l)_2} \ket{s_0}\ket{s'=0} \ket{(2^{2^n+l})_2}  \, ,   
\end{split}
\end{equation}
which completes the proof for $k\geq 2^n$.

\subsection{Parallelized scheme}

\begin{figure*}[t!]
\vspace{-1.5cm}
\hspace{-0.0cm}
	\includegraphics[width=3cm]{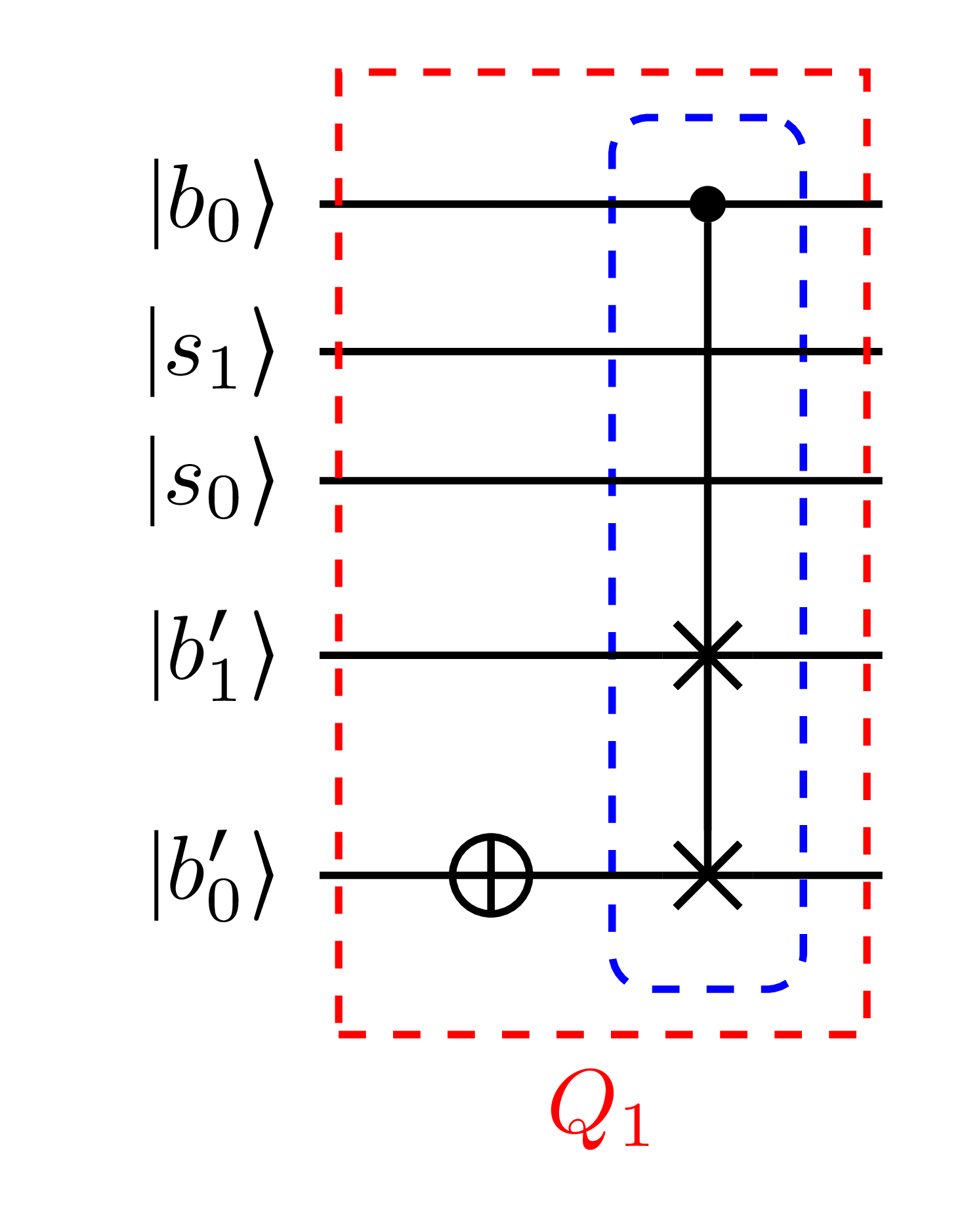}
	\includegraphics[width=5cm]{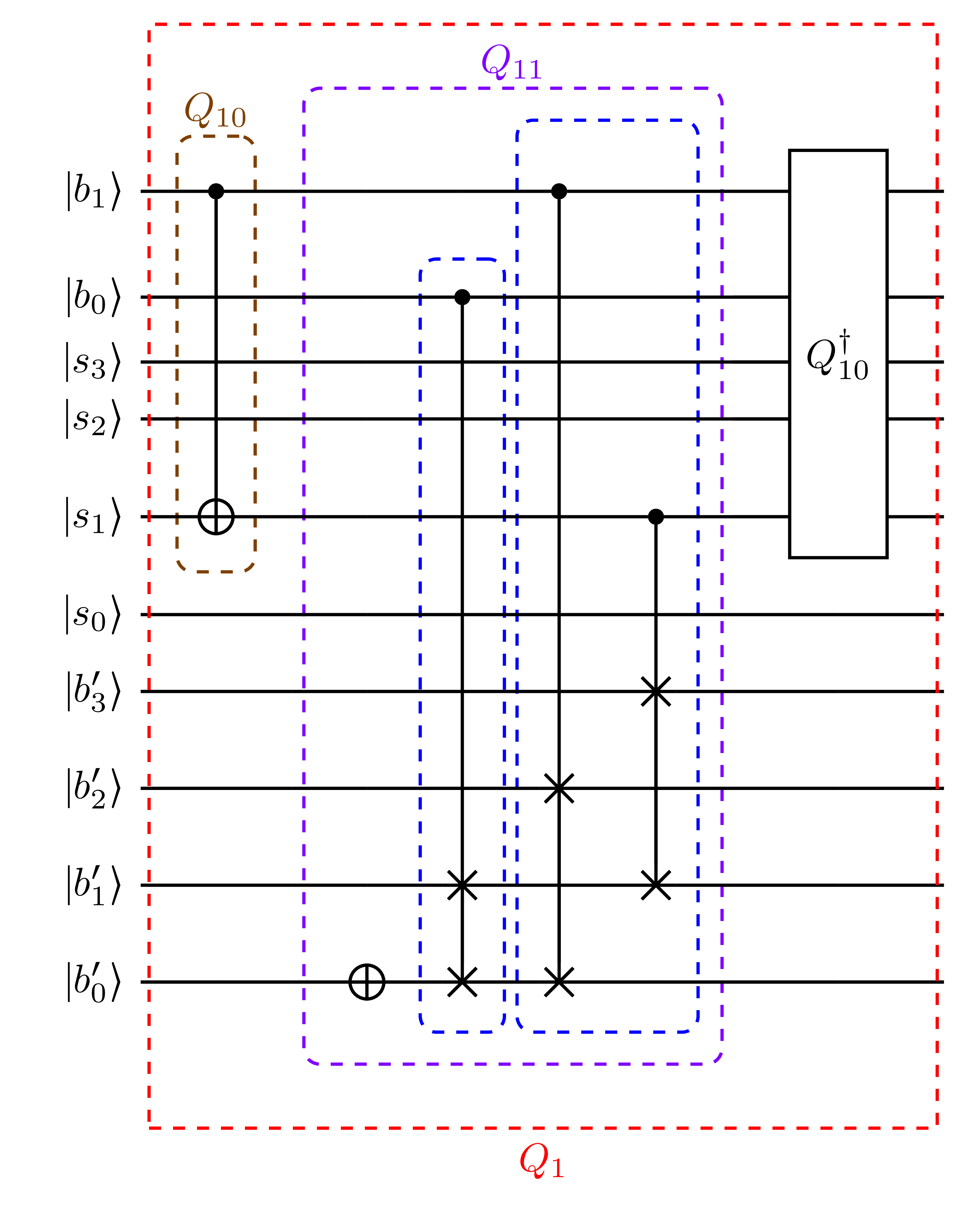}
	\includegraphics[width=7cm]{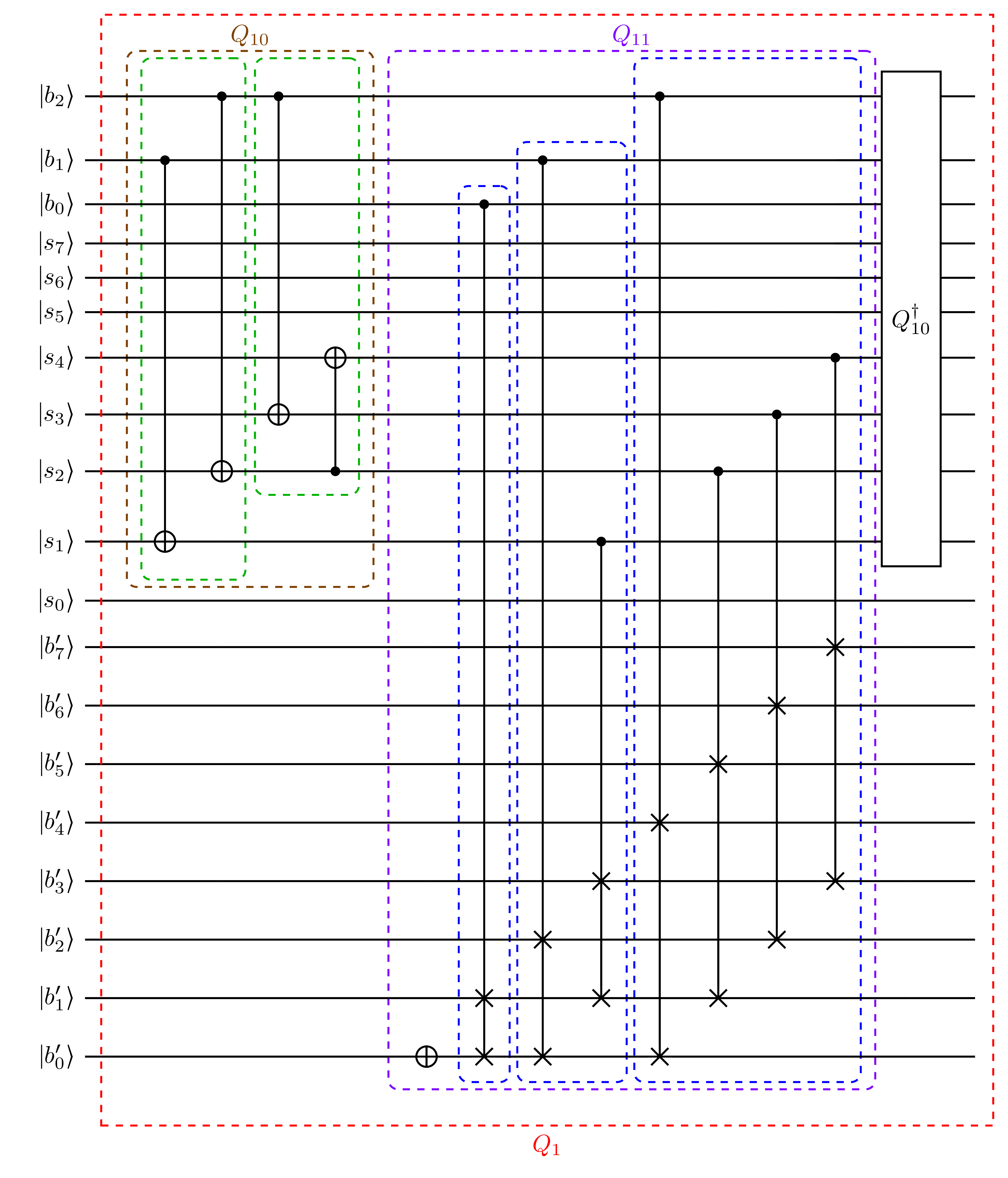}
	\vspace{-0.7cm}
    \includegraphics[width=16cm]{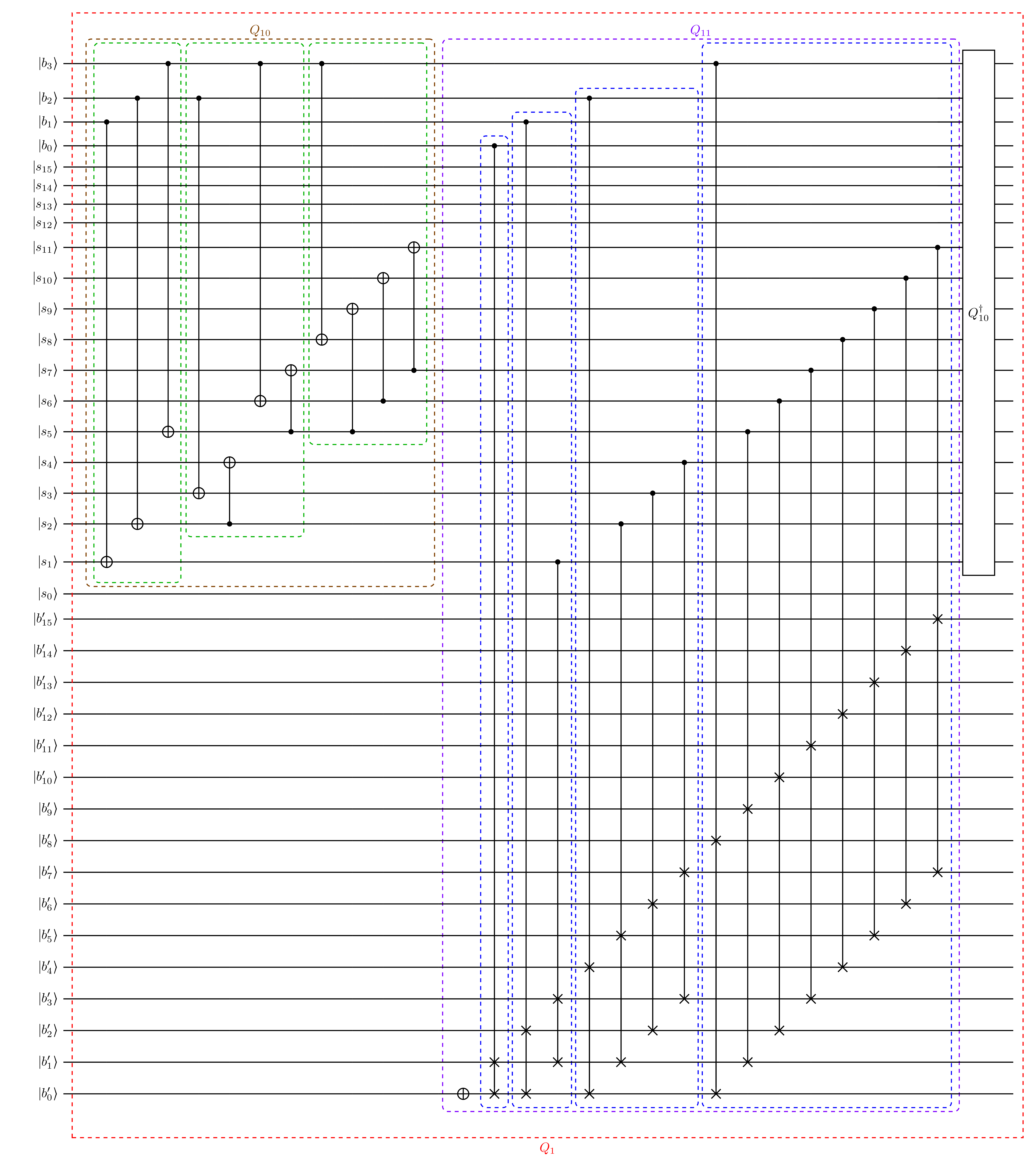}
    \vspace{-0.0cm}
	\caption{Diagrammatic representation of the parallelized scheme for $Q_1$ for $n=1$ (top left figure), $n=2$ (top middle figure), $n=3$ (top right figure), and $n=4$ (bottom figure). In this parallelized scheme, all operations inside a same blue or green box can be executed in parallel, which makes the total depth of the circuit linear in $n$. \label{fig:Q1parallelized}}
\end{figure*}

\begin{figure*}[t!]
\vspace{-0.0cm}
\hspace{-1.0cm}
	\includegraphics[width=19cm]{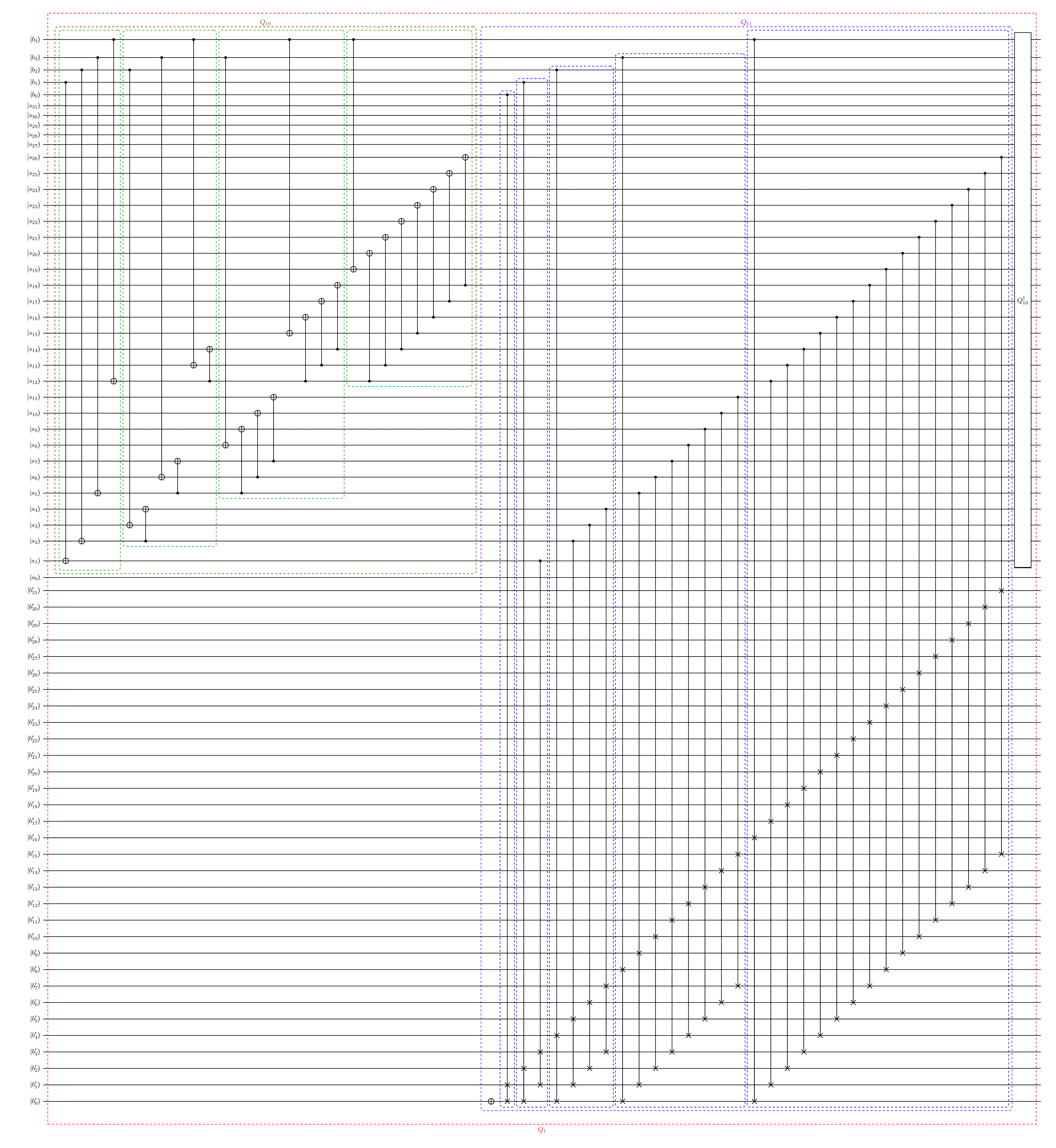}
    \vspace{-0.0cm}
	\caption{Diagrammatic representation of the parallelized scheme for $Q_1$, for $n=5$. In this parallelized scheme, all operations inside a same blue or green box can be executed in parallel, which makes the total depth of the circuit linear in $n$. \label{fig:Q1parallelizedbis}}
\end{figure*}

The circuit depicted in Fig.\ \ref{fig:Q1naive} and defined via Eqs.\ \eqref{eqs:Q1} implements a valid $Q_1$, but its depth is exponential in the number of position wires, i.e., in $n$, as it can be seen in Eq.\ \eqref{eq:Bnplus1}, since the product goes up to $i=2^n-1$ and all the SWAPs it implements are controlled on the same position qubit $b_{n-1}$.
We want to parallelize these SWAP operations.
For that, our idea is to copy the information contained in the position qubits onto the ancillary coins, which are available and in exponential number, and from the resulting ancillary coins we can initialize all ancillary positions at once as wished, since each SWAP can be controlled on a different ancillary coin, on which an appropriate copy of a position qubit has been done, after which we undo the copy operations performed on the ancillary coins.

Such a parallelized circuit is shown in Fig.\ \ref{fig:Q1parallelized} for $n=1$ to $4$, and in Fig.\ \ref{fig:Q1parallelizedbis} for $n=5$.

\subsubsection{Definition / Construction}

The parallelized circuit is defined as
\begin{equation}
Q_1 = Q_{10}^{\dag}  Q_{11} Q_{10}  \, ,
\end{equation}
where $Q_{10}$ corresponds to the operation of doing copies of the position qubits onto the ancillary coins, which we call ``copies operation'', and $Q_{11}$ corresponds to the parallelized controlled SWAPs. With $Q_{10}^{\dag}$ we undo the copies, so that after $Q_{10}^{\dag}$ all the ancillary coins $s_i$ are set back to $0$, as they were before applying $Q_{10}$.

The copies operation is performed with a series of CNOT gates applied appropriately; it is defined as follows,
\begin{align}
\label{eq:mainn}
Q_{10} \defeq \sideset{}{^L}\prod_{i=0}^{n-2} Q^{(i)}_{10} \, ,
\end{align}
where, for $i\geq 0$,
\begin{equation}
Q_{10}^{(i)} \defeq \sideset{}{^L}\prod_{m=i+1}^{n-1} J_m^{(i)} \, , 
\end{equation}
with
\begin{subequations}
\begin{align}
& \ \ \, J_m^{(0)} \defeq K_{b_m,s_{l_m}}(X) \\
&J_m^{(i\geq 1)} \defeq \nonumber \\
&\left( \sideset{}{^L}\prod_{l'=0}^{2^i-2} K_{s_{l_m+l'},s_{l_m+l'+2^i}}(X) \right) K_{b_m,s_{l_m+\sum_{u=1}^i 2^{u-1}}}(X) \, ,
\end{align}
\end{subequations}
and where we have defined
\begin{subequations}
\begin{align}
l_0 &\defeq 1  \\
l_m &\defeq 2^{m-1}-1 + l_{m-1} \ \ \text{for} \ m \geq 1 \, .
\end{align}
\end{subequations}

One can convince oneself that the copies operation (i) copies as many position qubits as needed on the ancillary coins, and (ii) is linear in $n$, since it doubles the number of copies performed at each incremented $i$ in Eq.\ \eqref{eq:mainn}.

It is easy to define $Q_{11}$ from Figs.\ \ref{fig:Q1parallelized} and \ref{fig:Q1parallelizedbis}: the difference with the naive scheme is that this time each SWAP is controlled on a \emph{different} ancillary coin (instead of on the same position qubit an exponential number of times), so that all these SWAPs can be applied simultaneously. \\

\section{Initializing the ancillary coins: $Q_2$}
\label{app:Q2}

\subsection{Basic requirement on $Q_2$}

The operator $Q_1$ acts as Eq.\ \eqref{eq:Q1action} on states $\ket K$,  so on states $\ket S$ of the form of Eq.\ \eqref{eq:special} it acts as
\begin{equation}
\label{eq:Q1S}
Q_1 \ket{S} = \sum_{k=0}^{2^n-1} \sum_{s_0=0,1} \alpha_{k,s_0} \ket{k_2} \ket{s_0} \ket{s'=0} \ket{(2^k)_2} \, .
\end{equation}
Equation \eqref{eq:Q1S} is the most general form of the state at the output of $Q_1$, and we want a unitary operator $Q_2$ which acts as follows on such a state:
\begin{widetext}
\begin{align}
Q_2 \left( Q_1 \ket{S} \right) &\equiv \sum_{k=0}^{2^n-1} Q_2 \left[ \ket{k_2} \left( \sum_{s_0=0,1} \alpha_{k,s_0} \ket{s_0} \right) \ket{s'=0} \ket{(2^k)_2}\right] \nonumber \\
&= \sum_{k=0}^{2^n-1} \ket{k_2} \ket{s_{2^n-1}=0} ... \ket{s_{k+1}=0} \! \left( \sum_{s_0=0,1} \alpha_{k,s_0} \ket{s_0} \right) \! \ket{s_{k-1}=0} ... \ket{s_0=0} \ket{(2^k)_2} \, .
\label{eq:outputQ2}
\end{align}
\end{widetext}
That is, $Q_2$ replaces the superposition that is on the $0$th coin by $\ket 0$, and puts this superposition on the $k$th coin instead of $\ket{s_{k}=0}$.

\subsection{Definition / Construction of $Q_2$}

\begin{figure*}[t!]
\vspace{-1cm}
\hspace{-0.5cm}
	\includegraphics[width=3cm]{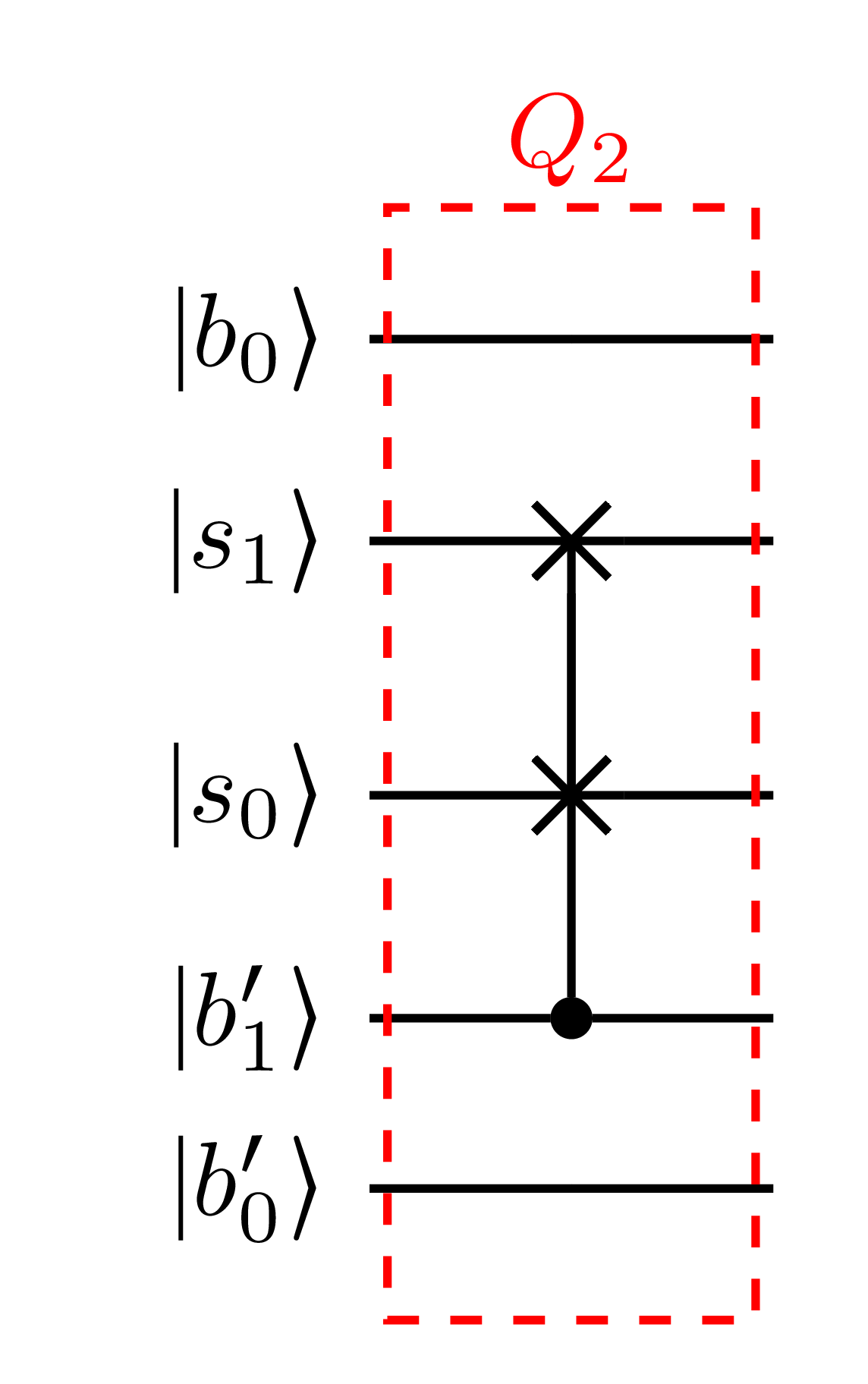}
	\includegraphics[width=4cm]{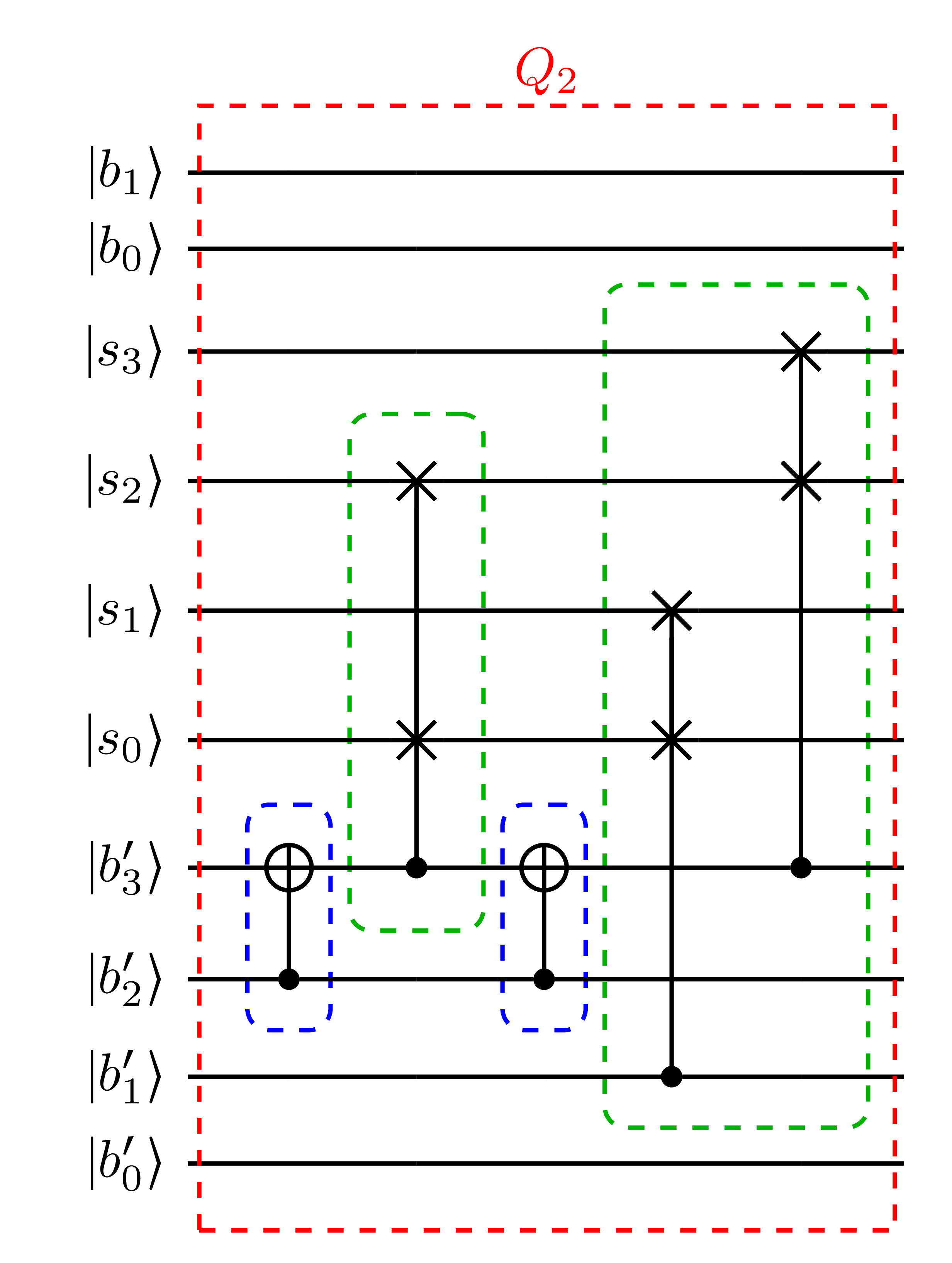}
	\includegraphics[width=6cm]{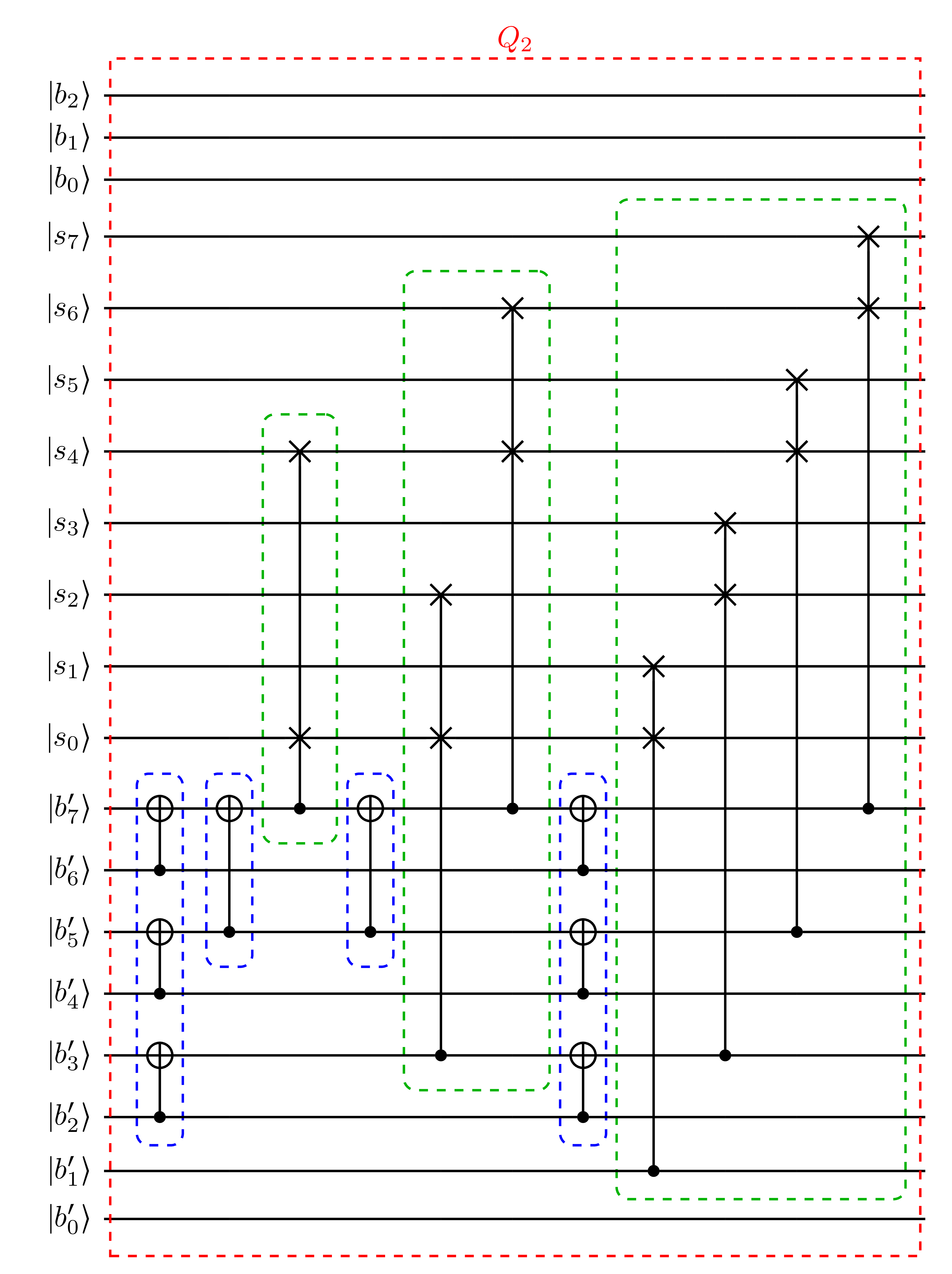}
	\includegraphics[width=12cm]{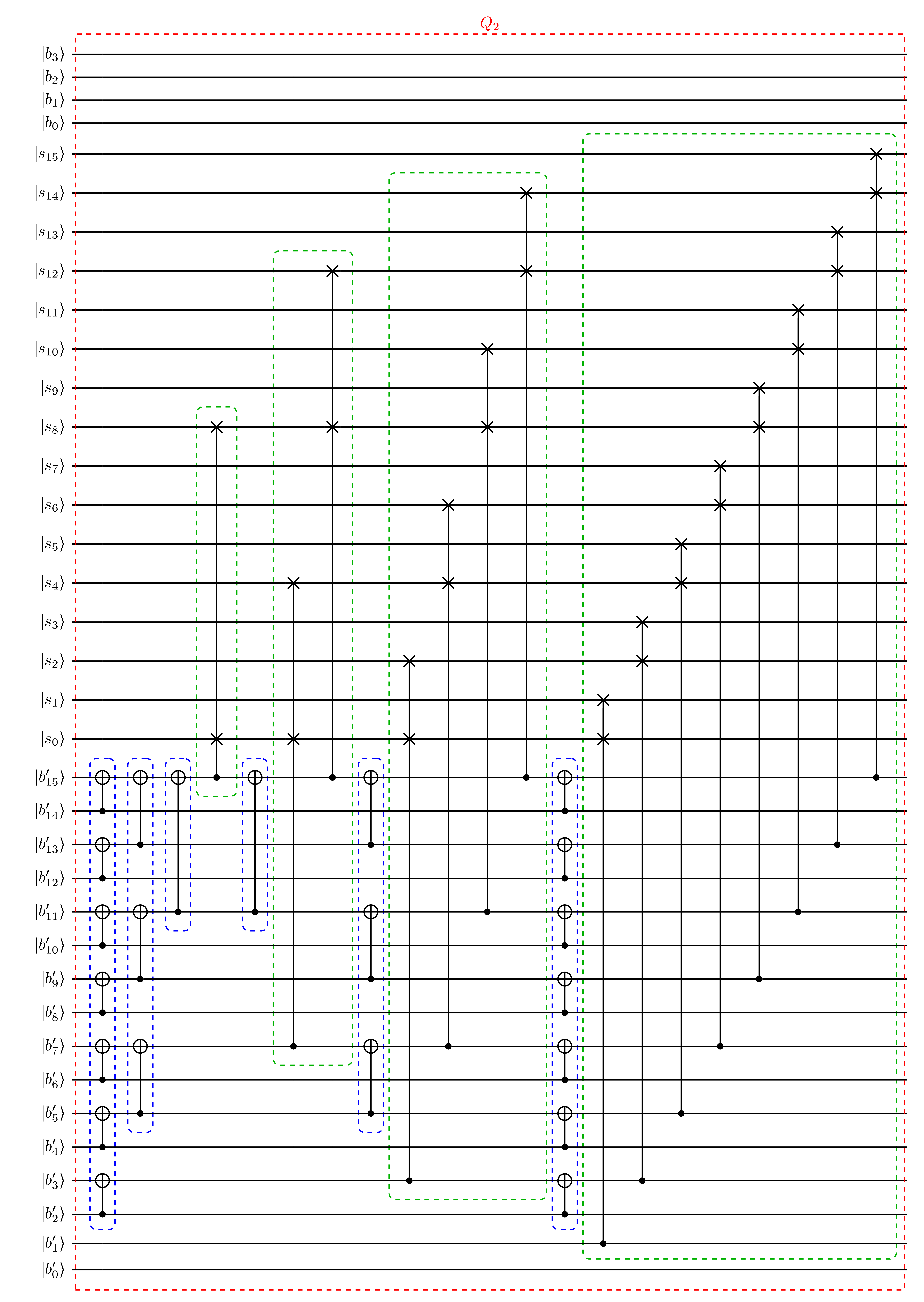}
	\vspace{-0.2cm}
	\caption{Diagrammatic representation of $Q_2$ for $n=1$ (top left figure), $n=2$ (top middle figure), $n=3$ (top right figure), and $n=4$ (bottom figure). All operations inside a same green box can be executed in parallel, which makes the total depth of the circuit linear in $n$. \label{fig:Q2}}
\end{figure*}

We want an operator $Q_2$ that is unitary and that satisfies Eq.\ \eqref{eq:outputQ2}; we call this a \emph{valid} $Q_2$.

We could use the position qubits in order to initialize the ancillary coins, similarly to how we have used them to initialize the ancillary positions, but then in order to parallelize the operations we would have to do copies as we have done to initialize the ancillary positions.
Instead, we are going to use the ancillary position qubits (which already contain all the information of the position qubits) to initialize the ancillary coins.

One can show that the circuit depicted in Fig.\ \ref{fig:Q2} for $n=1,2,3,4$, is a valid $Q_2$.
Let us briefly outline the proof, but before, let us define that circuit.

We use the notation
\begin{equation}
\mathbf{Q}^{(n)}_2 \defeq Q_2 \, ,
\end{equation}
which makes explicit the dependence in $n$.
We define $Q_2$ by induction as follows:
\begin{subequations}
\label{eqs:Q2}
\begin{align}
\mathbf{Q}^{(1)}_2 &\defeq E^{b'_1}_{s_0,s_1}
\label{eq:Q21} \\
\mathbf{Q}^{(n+1)}_2 &\defeq M^{(n+1)} V^{(n+1)} F(\mathbf{Q}^{(n)}_2) V^{(n+1)} \, ,
\end{align}
\end{subequations}
where
\begin{subequations}
\label{eqs:VM}
\begin{align}
V^{(n+1)} &\defeq \bigotimes_{m=1}^{2^n-1} K_{b'_{2m},b'_{2m+1}}(X) \\
M^{(n+1)} &\defeq \bigotimes_{l=0}^{2^n-1} E^{b'_{1+2l}}_{s_{2l},s_{1+2l}} \, ,
\end{align}
\end{subequations}
and where $F(.)$ takes the circuit $\mathbf{Q}^{(n)}_2$ and multiplies by 2 all ``heights'', that is, it performs a vertical homothety, or dilation, of factor 2, with center on $s_0$ and $b'_{2^n-1}$.
In Eqs.\ \eqref{eq:Q21} and \eqref{eqs:VM}, we have purposely omitted the various tensor factors $I_2$ on all remaining qubits, in order to lighten notations.

Let us explain a bit how $\mathbf{Q}^{(n+1)}_2$ works, i.e., let us outline the proof to show that the circuit defined via Eqs.\ \eqref{eqs:Q2} is a valid $Q_2$.
If we apply solely $F(\mathbf{Q}^{(n)}_2)$ to an input state for which the auxiliary position  $b'_k$ that is ``turned on'' (i.e., the value of which is $1$) is even, i.e., $k=2p$ with $p$ an integer, no operation is performed.
If we now apply this $F(\mathbf{Q}^{(n)}_2)$ to an input state for which $b'_k=1$ for $k$ odd, then the auxiliary coins which are swapped are $s_0$ and $s_{k-1}$ (and recall that we want $s_k$ instead of $s_{k-1}$).
Thanks to $V^{(n+1)} \cdot V^{(n+1)}$, applying  $V^{(n+1)} F(\mathbf{Q}^{(n)}_2) V^{(n+1)}$ to an input state for which $b'_k=1$ for $k$ even does swap $s_0$ and $s_k$, without the need of $M^{(n+1)}$.
What $M^{(n+1)}$ does is to have the auxiliary coin that is exchanged with $s_0$ when $b'_k=1$ for $k$ odd, pass from $s_{k-1}$ to $s_k$; $M^{(n+1)}$ only acts on odd auxiliary positions.

The depth of both $V^{(n+1)}$ and $M^{(n+1)}$ is $1$, and that of $F(\mathbf{Q}^{(n)}_2) $ is the same as that of $\mathbf{Q}^{(n)}_2 $, so that the depth of $Q_2$ is linear in $n$. \\


\section{Proof that $U^{(n)}_{\mathrm{lin.}}$ coincides with ${C}^{(n)}$ on all relevant states}
\label{app:proof_linear}

On the one hand, the state at the output of $Q_2$ has the form of Eq.\ \eqref{eq:outputQ2}.
On the other hand, $Q_0$ is defined by Eq.\ \eqref{eq:Q0}.
So, $Q_0$ acting on $Q_2Q_1 \ket{S}$ gives
\begin{widetext}
\begin{equation}
\begin{split}
   Q_0Q_2Q_1 \ket{S} 
= \sum_{k=0}^{2^n-1} \ket{k_2} \ket{s_{2^n-1}=0} ... \ket{s_{k+1}=0} \left( C_k \sum_{s_0=0,1} \alpha_{k,s_0} \ket{s_0} \right)  \ket{s_{k-1}=0} ... \ket{s_0=0} \ket{(2^k)_2} 
\end{split}
\end{equation}
We know from Eq.\ \eqref{eq:outputQ2} how $Q_2^{-1}$ acts on states of the form of the right-hand side of Eq.\ \eqref{eq:outputQ2}, that is, states that can be written under the form $Q_2 Q_1 \ket{S}$.
Now, because each $C_k$ is unitary, it is easy to prove that the state $Q_0 Q_2 Q_1 \ket{S}$ is also of the form of the right-hand side of Eq.\ \eqref{eq:outputQ2}, i.e., it is a state of the form 
\begin{equation}
   Q_0Q_2Q_1 \ket{S} 
= \sum_{k=0}^{2^n-1} \ket{k_2} \ket{s_{2^n-1}=0} ... \ket{s_{k+1}=0} \left( \sum_{s_0=0,1} \tilde{\alpha}_{k,s_0} \ket{s_0} \right)  \ket{s_{k-1}=0} ... \ket{s_0=0} \ket{(2^k)_2} \, ,
\end{equation}
with $\sum_{k=0}^{2^n-1} \sum_{s_0=0,1} |\tilde{\alpha}_{k,s_0}|^2 = \sum_{k=0}^{2^n-1} \sum_{s_0=0,1} |{\alpha}_{k,s_0}|^2 =  1$.
Hence, we know that 
\begin{subequations}
\begin{align}
Q_2^{-1} Q_0Q_2Q_1 \ket{S} 
&=  \sum_{k=0}^{2^n-1} \ket{k_2} \left( \sum_{s_0=0,1} \tilde{\alpha}_{k,s_0} \ket{s_0} \right) \ket{s'=0} \ket{(2^k)_2} \\
&= \sum_{k=0}^{2^n-1}  \sum_{s_0=0,1}  \tilde{\alpha}_{k,s_0} \ket{k_2}  \ket{s_0} \ket{s'=0} \ket{(2^k)_2} \, .
\label{eq:finalll}
\end{align}
\end{subequations}
We know from Eq.\ \eqref{eq:Q1S} how $Q_1^{-1}$ acts on states of the form of Eq.\ \eqref{eq:finalll} (i.e., of the form $Q_1 \ket{S}$).
Hence, applying $Q_1^{-1}$ to $Q_2^{-1} Q_0Q_2Q_1 \ket{S} $ given by Eq.\ \eqref{eq:finalll} yields
\begin{subequations}
\begin{align}
 Q_1^{-1} Q_2^{-1} Q_0 Q_2 Q_1 \ket{S} &=  \sum_{k=0}^{2^n-1}  \sum_{s_0=0,1}  \tilde{\alpha}_{k,s_0} \ket{k_2}  \ket{s_0} \ket{s'=0} \ket{b'=0}  \\
 &= \sum_{k=0}^{2^n-1} \ket{k_2} \left( C_k \sum_{s_0=0,1} \alpha_{k,s_0} \ket{s_0} \right) \ket{s'=0} \ket{b'=0} \\
 &=  \sum_{k=0}^{2^n-1}  \sum_{s_0=0,1}  \alpha_{k,s_0} \ket{k_2}  (C_k  \ket{s_0})  \ket{s'=0} \ket{b'=0} \\
 &=  \left[ \left( \sum_{k'=0}^{2^n-1} \ket{k'_2} \! \bra{k'_2} \otimes C_k \right) \sum_{k=0}^{2^n-1} \sum_{s_0=0,1}  \alpha_{k,s_0} \ket{k_2}  \ket{s_0} \right]  \ket{s'=0} \ket{b'=0}  \\
&= \left( {C}^{(n)} \sum_{k=0}^{2^n-1}  \sum_{s_0=0,1}  \alpha_{k,s_0} \ket{k_2}  \ket{s_0} \right)  \ket{s'=0} \ket{b'=0} \\
&= \left( {C}^{(n)} \otimes I_{2^{(2^n-1)}} \otimes I_{2^{(2^n)}} \right)  \left( \sum_{k=0}^{2^n-1}  \sum_{s_0=0,1}  \alpha_{k,s_0} \ket{k_2}  \ket{s_0}  \ket{s'=0} \ket{b'=0}\right) \\
&= \left( {C}^{(n)} \otimes I_{2^{(2^n-1)}} \otimes I_{2^{(2^n)}} \right) \ket{S} \, .
\label{eq:Ffinal}
\end{align}
\end{subequations}
\end{widetext}
The right-hand side of Eq.\ \eqref{eq:Ffinal} is that of Eq.\ \eqref{eq:the_thing}, and the left-hand side of Eq.\ \eqref{eq:Ffinal} is also that of Eq.\ \eqref{eq:the_thing} because  $Q_1^{-1} = Q_1^{\dag}$, $Q_2^{-1} = Q_2^{\dag}$, and $U_{\text{lin.}}^{(n)}$ is defined by Eq.\ \eqref{eq:U_linear}, which completes the proof.

\section{Efficient quantum circuit for a smooth position-dependent coin operator}
\label{app:EQC}

\subsection{The exact, complete circuit, for a complete Walsh decomposition}
\label{subapp:the_circuit}

We follow the same computational steps as in Ref.\ \cite{WGMAG2014}. 
In this subappendix, we first derive an exact quantum circuit for $C^{(n)}_{f,\sigma}$ using the development of $f$ into a Walsh series.  

\subsubsection{$C^{(n)}_{f,\sigma}$ as a product of Walsh terms $U_j$}

First, let us define the Walsh functions: for any natural number $j$ and real number $x \in [0,1]$, the $j$th Walsh function at point $x$ is
\begin{equation}
    w_j(x)\defeq(-1)^{\sum_{i=1}^nj_ix_{i-1}} \, ,
\end{equation}
where $j_i$ is the $i$th bit in the binary expansion $j=\sum_{i=1}^nj_i2^{i-1}$, with $n$ the most significant non-vanishing bit of $j$, and $x_i$ the $i$th bit in the dyadic expansion\footnote{Here the sum of the dyadic expansion is up to $\infty$ because we expand \emph{any} real number in $[0,1]$; the sum is finite only when we encode powers of $2$.} $x=\sum_{i=0}^{\infty} x_i/ 2^{i+1}$.
The Walsh functions form a complete set of orthonormal functions.
For a finite number of points $x_p$ one can define the discrete Walsh functions $w_{jp} \defeq w_j(x_p)$.
The set of discrete Walsh functions is still complete and orthormal: $\frac{1}{2^n}\sum_{p=0}^{2^n-1}w_{jp}w_{lp}=\delta_{jl}$ and $\frac{1}{2^n}\sum_{j=0}^{2^n-1}w_{jp}w_{jl}=\delta_{pl}$.
Thus, one can expand the $f(x_p)$'s in terms of discrete Walsh functions as
\begin{equation}
    f(x_p)=\sum_{j=0}^{2^n-1} a_j w_{jp} \, ,
\end{equation}
where the $a_j$ are real numbers such that
\begin{equation}
    a_j\defeq \frac{1}{2^n}\sum_{p=0}^{2^n-1}f(x_p)w_{jp} \, .
\end{equation}

The Walsh functions at a given position can only take the value $+1$ or $-1$, making their associated Walsh operators a tensor product of $Z$ Pauli operators.
Indeed, we define the Walsh operators as
\begin{equation}
\label{eq:Walsh_op}
    \hat{w}_j=(Z_{n-1})^{j_n}\otimes\cdots\otimes(Z_0)^{j_1} \, ,
\end{equation}
where $Z_i$ is a third Pauli matrix applied on qubit $i$, and $j=0,\cdots,2^n-1$.
Therefore, for each real number $x_p=\sum_{i=0}^{n-1}b_i/2^{i+1}$, with $b_i\in \{0,1\}$, one has
\begin{subequations}
\begin{align}
\hat{w}_j\ket{k_2} &\equiv \hat{w}_j\ket{b_{n-1}\cdots b_0} \\
&= (-1)^{\sum_{i=1}^nj_ib_{i-1}}\ket{b_{n-1}\cdots b_0} \\
&= w_{jp}\ket{k_2} \, .
\end{align}
\end{subequations}

The operator $C^{(n)}_{f,\sigma}$, defined in Eq.\ \eqref{eq:Cnfsigma}, can now be rewritten in terms of Walsh operators:
\begin{subequations}
\label{eqs:blabla}
\begin{align}
    C^{(n)}_{f,\sigma} & \defeq e^{i\hat{f}\otimes \sigma} \\
    &=e^{i\sum_{j=0}^{2^n-1}a_j\hat{w}_j \otimes \sigma} \\
    &=\prod_{j=0}^{2^n-1} U_j \, ,
\end{align}
\end{subequations}
where
\begin{equation}
U_j \defeq  e^{ia_j\hat{w}_j\otimes \sigma} \, ,
\end{equation}
and where, in going from the second to the third line in Eqs. \eqref{eqs:blabla}, we have used the following commutation relations, $\forall j,j'$, $ [\hat{w}_j\otimes \sigma ,\hat{w}_{j'} \otimes \sigma ]=0$.

\subsubsection{Quantum circuits for the Walsh terms $U_j$}

Let us now derive the quantum circuits for the operators $U_j$.
Let us consider, in the expression the Walsh operators $\hat w_j$, Eq.\ \eqref{eq:Walsh_op}, only the $Z$ Pauli matrices on the relevant qubits (i.e., one for each $j_i=1$), and omit the other, $I_2$ tensor factors (i.e., one for each $j_i=0$).
Let $r$ be the number of such $Z$ Pauli matrices.
Now, we first remark that every tensor product of a number $r$ of $Z$ Pauli matrices can be rewritten using CNOT matrices:
\begin{equation}
    Z_{r-1}\otimes \cdots \otimes Z_1 \otimes Z_0 = A_r(I_{2^{(r-1)}}\otimes...\otimes I_{2^{(1)}}\otimes Z_0)A_r^{-1} \, ,
\end{equation}
where $A_r \defeq CNOT_{1}^{0} \cdot CNOT_{2}^{0}...CNOT_{r-1}^{0}$ and $CNOT_i^j$ is the CNOT quantum gate controlled by qubit $i$ and applied on qubit $j$.
Therefore, the operator $U_j$ written as acting on the $r$ relevant position qubits and the coin qubit, can be written in terms of quantum gates as
\begin{equation}
    U_j=(A_r\otimes I_{2})(I_{2^{(r-1)}}\otimes...\otimes I_{2^{(1)}}\otimes e^{i a_jZ\otimes \sigma})(A_r^{-1}\otimes I_{2}) \, .
\end{equation}

\subsubsection{Quantum circuits for the terms $e^{i a_jZ\otimes \sigma}$}

Let us finally derive a quantum circuit for each $e^{i a_jZ\otimes \sigma}$, for $\sigma=X,Y,Z$. 
For $\sigma=Z$, the quantum circuit is simply
\begin{subequations}
\begin{align}
e^{i a_jZ\otimes Z}&=CNOT(I_{2}\otimes e^{ia_jZ})CNOT \\
&=CNOT(I_{2}\otimes R_Z(-2a_j))CNOT \, ,
\end{align}
\end{subequations}
where $R_Z(\theta)=\begin{bmatrix} e^{-i\frac{\theta}{2}}&0 \\ 0&e^{i\frac{\theta}{2}}\end{bmatrix}$.

For $\sigma=Y$:
\begin{subequations}
\begin{align}
    e^{ia_jZ\otimes Y} &=
    \begin{bmatrix} 
    e^{ia_jY} & 0 \\
    0 & e^{-ia_jY}
    \end{bmatrix} \\
    &= \begin{bmatrix} 
    e^{ia_jY} & 0 \\
    0 & e^{ia_jXYX}
    \end{bmatrix} \\
    &= \begin{bmatrix} 
    e^{ia_jY} & 0 \\
    0 & Xe^{ia_jY}X
    \end{bmatrix} \\
    &= CNOT(I_{2}\otimes e^{ia_jY})CNOT \\
    &=CNOT(I_{2}\otimes R_Y(-2a_j))CNOT \, ,
\end{align}
\end{subequations}
where $R_Y(\theta)=\begin{bmatrix} \cos (\frac{\theta}{2})&-\sin(\frac{\theta}{2}) \\ \sin(\frac{\theta}{2})&\cos (\frac{\theta}{2}) \end{bmatrix}$.

For $\sigma=X$:
\begin{subequations}
\begin{align}
    e^{ia_jZ\otimes X} &=
    \begin{bmatrix} 
    e^{ia_jX} & 0 \\
    0 & e^{-ia_jX}
    \end{bmatrix} \\
    &= \begin{bmatrix} 
    e^{ia_jY} & 0 \\
    0 & e^{ia_jZXZ}
    \end{bmatrix} \\
    &= \begin{bmatrix} 
    e^{ia_jX} & 0 \\
    0 & Ze^{ia_jX}Z
    \end{bmatrix} \\
    &= \widehat{CZ}(I_{2}\otimes e^{ia_jX})\widehat{CZ} \\
    &= \widehat{CZ}(I_{2}\otimes R_X(-2a_j))\widehat{CZ} \, ,
\end{align}
\end{subequations}
where $R_X(\theta)\defeq\begin{bmatrix} \cos (\frac{\theta}{2})&-i\sin(\frac{\theta}{2}) \\ -i\sin(\frac{\theta}{2})&\cos (\frac{\theta}{2}) \end{bmatrix}$ and $\widehat{CZ}\defeq\text{diag}(1,1,1,-1)$ is the controlled-$Z$ gate.

Since the operators $U_j$ commute with each other, the implementation of $C^{(n)}_{f_m,\sigma}=\prod_{j=0}^{2^m-1}U_j$ can be optimized by choosing the right sequence that cancels a maximum number of CNOT gates.
This can be done by using the Gray code \cite{book_beauchamp1984applications}, resulting into only one CNOT gate per Walsh operator implemented and improving polynomialy the complexity of the scheme.
Such an optimization could perhaps also be done by the transpiler of the used quantum computer.

We have finally obtained an exact quantum circuit $C^{(n)}_{f,\sigma}=\prod_{j=0}^{2^n-1}U_j$, where the order of the terms in the product is determined by the previous optimization.

\subsection{Approximation by truncation of the complete circuit, and related efficiency}
\label{subapp:efficiency}

\subsubsection{Idea}

The idea here is the following.
Instead of implementing the exact, complete decomposition $C^{(n)}_{f,\sigma}=\prod_{j=0}^{2^n-1}U_j$, the depth of which scales exponentially with $n$, we truncate the circuit up to some $j=2^m$, i.e., we approximate $C^{(n)}_{f,\sigma}$ by some truncation $C^{(n)}_{f_m,\sigma} = \prod_{j=0}^{2^m-1}U_j $, where we have introduced the truncation $f_m\defeq \sum_{j=0}^{2^m-1} a_j w_{j} $ of the Walsh development of a function $f$ up to the term $2^m$.
This is interesting only if $m$ does not scale exponentially with $n$.
Notice that even if we truncate the \emph{quantum circuit} $C^{(n)}_{f,\sigma}$ into $C^{(n)}_{f_m,\sigma}$, we still need the full decomposition of the \emph{function} $f$ in order to compute accurately the coefficients $a_j$ of the decomposition. \\

\subsubsection{Approximation error and smoothness condition}

Let $\epsilon$ be a positive real number.
One can show that $\sup_x|f(x)-f_m(x)|<\frac{\sup_{x \in [0,1]} f'(x)}{2^m}$ \cite{yuen1975function, book_golubov2012walsh}, where $x$ is a continuous or discrete variable in $[0,1]$.
This implies that for $m$ such that $\frac{\sup_{x \in [0,1]} f'(x)}{2^m} = \epsilon$, the approximating coin operator $C^{(n)}_{f_m,\sigma}$ is close to the real coin operator $C^{(n)}_{f,\sigma}$ up to $\epsilon$ in terms of the spectral norm.
Therefore, only $2^m  \propto 1/\epsilon $ Walsh operators have to be implemented to approximate the coin operator up to $\epsilon$, resulting in a quantum circuit with a total number of one-qubit and two-qubit quantum gates of $O(1/\epsilon)$.
The condition for the quantum circuit to be efficient is $m\ll n$, i.e., $2^m \ll 2^n$, i.e., $\epsilon 2^m \ll \epsilon 2^n$, which, using the fact that $\sup_{x \in [0,1]} f'(x) = \epsilon 2^m$, yields Eq.\ \eqref{eq:smoothness_cond}, which is a smoothness condition on $f$.

\subsection{Linear position dependence}
\label{subapp:linear_pos_dep}

\begin{figure*}
\begin{quantikz}
\lstick{$\ket{x_{n-1}}$} & \ctrl{4} & \qw & [2cm] \qw & \qw  \\ 
 \lstick{$\ket{x_{n-2}}$}&\qw & \ctrl{3} & \qw & \qw  \\
& \vdots & & & & \\
 \lstick{$\ket{x_{0}}$}&\qw & \qw& \ctrl{1}  & \qw \\
 \lstick{$\ket{\phi}=\alpha\ket{0}+\beta\ket{1}$}&\gate{R(a/2^{n})} & \gate{R(a/2^{n-1})} & \gate{R(a/2)}& \qw
\end{quantikz} 
\caption{Circuit implementing the block-diagonal unitary matrix $C^{(n)}_{f,\sigma}\defeq e^{i\hat{f}\otimes\sigma}$ for $f$ being a linear function $f(x) \defeq ax$. This circuit only uses $n$ two-qubit gates.\label{fig:linear_function}}
\end{figure*}
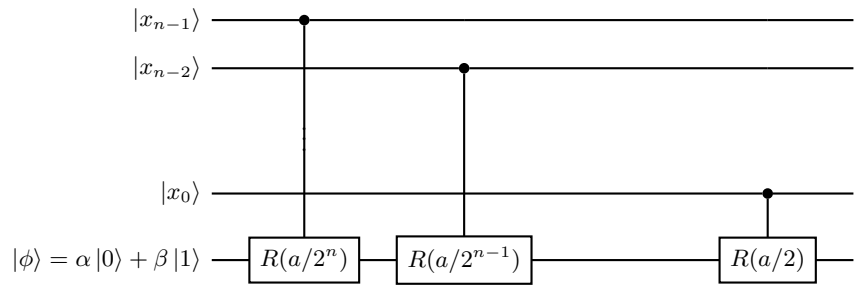

In the particular case of a unitary operator $C^{(n)}_{f,\sigma}\defeq e^{i\hat{f}\otimes\sigma}$ with $\hat{f}$ the operator associated to a function $f$ that is \emph{linear}, i.e., $f(x) \defeq ax$ where $x \in [0,1]$ and $a \in \mathbb{R}$, one can derive a quantum circuit using only $n$ two-qubit gates.
Indeed, let us consider the quantum circuit of Fig.\ \ref{fig:linear_function} acting on the state $\ket{\psi}=\ket{x_{n-1}}\cdots \ket{x_0}\ket{\phi}$, and let $R(\theta) \defeq e^{i\theta\sigma}$.
Each rotation controlled by $\ket{x_i}$ can be written as $R(x_i\theta /2^{i+1})$ where $x_i=0,1$.
The final state can be written as
\begin{subequations}
\begin{align}
    \ket{\psi} &=\ket{x_{n-1}}\cdots \ket{x_{0}}R(a\sum_{i=0}^{n-1}x_i/2^{i+1})\ket{\phi} \\
    &=\ket{x_{n-1}}\cdots \ket{x_{0}}R(ax)\ket{\phi} \, ,
\end{align}
\end{subequations}
where $x=\sum_{i=0}^{n-1}x_i/2^{i+1}$ is given by its dyatic expansion.
This quantum circuit applied on a general state $\ket{\psi}_i=\sum_x \ket{x}(c_x^0\ket{0}+c_x^1\ket{1})$ will give the final state:
\begin{subequations}
\begin{align}
    \ket{\psi}_f&=\sum_x c_x\ket{x} R(ax)(c_x^0\ket{0}+c_x^1\ket{1}) \\
&=e^{i\hat{f}\otimes \sigma}\ket{\psi}_i \, .
\end{align}
\end{subequations}

\end{document}